%% file: main.tex
\documentclass[sigplan,nonacm]{acmart}
\PassOptionsToPackage{font={small}}{caption}
\usepackage{multirow}
\usepackage{array}
\usepackage{microtype}

\usepackage{soul}
\usepackage[normalem]{ulem}

\def\pname{LSM-GNN}

\newcommand{\bp}[1]{{\textcolor{orange}{Brian: \textit{#1}}}}

\newcommand{\vm}[1]{{\textcolor{blue}{Vikram: \textit{#1}}}}
\newcommand{\zq}[1]{{\textcolor{blue}{Zaid}}}

\begin{document}

\title{
LSM-GNN: Large-scale Storage-based Multi-GPU GNN Training by Optimizing Data Transfer Scheme
}

\author{Jeongmin Brian Park}
\affiliation{%
  \institution{UIUC}\country{USA}
  }
\email{jpark346@illinois.edu}

\author{Kun Wu}
\affiliation{%
  \institution{UIUC}\country{USA}
  }
\email{kunwu2@illinois.edu}

\author{Vikram Sharma Mailthody}
\affiliation{%
  \institution{NVIDIA}\country{USA}
  }
\email{vmailthody@nvidia.com}

\author{Zaid Qureshi} 
\affiliation{%
  \institution{NVIDIA}\country{USA}
  }
\email{zqureshi@nvidia.com}

\author{Scott Mahlke}
\affiliation{%
  \institution{NVIDIA}\country{USA}
  }
\email{smahlke@nvidia.com}

\author{Wen-mei Hwu}
\affiliation{%
  \institution{NVIDIA/UIUC}\country{USA}
  }
\email{whwu@nvidia.com}

\input{abstract}

\maketitle
\pagestyle{plain}

\thispagestyle{empty}





\input{introduction}

\input{background}

\input{motivation}

\input{design}

\input{Evaluation}

\input{related}

\input{conclusion}

\bibliographystyle{plain}
\bibliography{Reference/references,Reference/_GENERATED_from_Kuns_Zotero,Reference/GNN_backgrounds_Kun_manually_added}

\end{document}

%% file: abstract.tex
\begin{abstract}

Graph Neural Networks (GNNs) are widely used today in recommendation systems, fraud detection, and node/link classification tasks. 
Real world GNNs continue to scale in size and require a large memory footprint for storing graphs and embeddings that often exceed the memory capacities of the target GPUs used for training. 
To address limited memory capacities, traditional GNN training approaches use graph partitioning and sharding techniques to scale up across multiple GPUs within a node and/or scale out across multiple nodes.  
However, this approach suffers from the high computational costs of graph partitioning algorithms and inefficient communication across GPUs. 




To address these overheads, we propose Large-scale Storage-based Multi-GPU GNN framework (\pname{}), a storage-based approach to train GNN models that 
utilizes a novel communication layer enabling GPU software caches to function as a system-wide shared cache with low overheads. 
\pname{} incorporates a hybrid eviction policy that intelligently manages cache space by using both static and dynamic node information to significantly enhance cache performance.
Furthermore, we introduce the Preemptive Victim-buffer Prefetcher (PVP), a mechanism for prefetching node feature data from a Victim Buffer located in  CPU pinned-memory to further reduce the pressure on the storage devices. 
Experimental results show that despite the lower compute capabilities and memory
capacities, LSM-GNN in a single node with two GPUs offer superior performance over two-node-four-GPU Dist-DGL baseline and provides up to 3.75$\times$ speed up on end-to-end epoch time while running large-scale GNN training.


\end{abstract}

%% file: introduction.tex
\section{INTRODUCTION}

Graph neural networks (GNNs) have emerged as an effective paradigm for learning rich relation and interaction information among input nodes and edges, leading to improved generalization performance over traditional machine learning techniques.
Consequently, GNNs have gained significant traction in recent years and demonstrated their efficacy in 
graph-based machine learning applications, such as recommendation~\cite{gnn-recommendation, pinner_sage}, fraud detection~\cite{gnn_fraud,fdgar,gnn_fraud3,gnn_fraud4}, node classification~\cite{GCN,Graphsage,GAT,LazyGCN}, and link prediction~\cite{gnn_linkpredict,fewshot,LP_system}.

To cater to this growing interest, new open-source frameworks such as PyTorch Geometric (PyG)~\cite{pyg}, Spektral~\cite{Spektral}, and Deep Graph Library (DGL)~\cite{dgl} have been developed to provide performant GNN execution. 
To increase performance, they incorporate optimizations, such as message-passing for aggregating feature information across related graph nodes, and graph-specific neural network computation layers. These optimizations improve the speed of all stages of  GNN training (sampling, feature aggregation, and model training) when the entire graph and embedding data can fit within a single GPU's memory.


The challenge escalates with real-world graphs that can scale to massive proportions, far exceeding the memory capacities of traditional systems. For example, the Pinterest user-to-item graph contains over 2 billion nodes and 17 billion edges, totaling a data size of 18 TB~\cite{pinner_sage}. 
Such sizes render it impractical to load the entire graph and embeddings into the memory of one GPU or even the host CPU.  
Sharding the graph and embeddings across many CPU and/or GPU memories limits scalability. 
In a multi-GPU environment, scalability is limited because each GPU needs to access most of its required features from other GPU's or CPU's memories, resulting in high network communication overhead.

To reduce communication overhead,  state-of-the-art distributed GNN training often partitions the graph with well-known graph partition algorithms like METIS ~\cite{Metis}. 
However, graph partitioning algorithms are notoriously time-consuming and their execution time increases exponentially with graph size resulting in significant preprocessing overhead that must be amortized over the training time. 
Furthermore, the graph partitioning algorithm consumes a significant amount of memory for temporary data and can run out of memory while processing large graphs even on high-memory capacity CPUs.  
For example, when partitioning a medium-sized heterogeneous IGBH-medium~\cite{IGB} graph in DistDGL~\cite{distdgl}, it takes more than 220 GB of memory, surpassing 5 times the size of the dataset. On a node with 4$\times$ AMD ``Interlagos'' CPUs, partitioning IGBH-medium into four partitions via METIS  takes more than 5 times the time of random partitioning, as detailed in Section~\ref{sec:bg_dist_training}.

An alternative approach is to use storage-based GNN training, where prior works~\cite{GIDS, Ginex, MariusGNN} maintain graph data in storage and feed sampled data to the GPU on demand. 
This method is cost-effective and obviates the need for extensive additional resources dedicated to GNN training. However, the feature aggregation stage is predominantly constrained by the bandwidth of storage. 
To effectively scale storage-based GNN training in multi-GPU environments, a proportional increase in the number of SSDs is necessary to ensure adequate storage bandwidth for each GPU. However, it may not be practical to increase the number of SSDs in many systems.
Moreover, altering the SSD configurations to accommodate system changes necessitates either data replication or reformatting. 
Such operations introduce substantial overhead, especially for large-scale graphs. 
Thus, a naive adaption of the storage-based GNN training to multi-GPU systems is not practical.

To effectively support storage-based multi-GPU training, we introduce the Large-scale Storage-based Multi-GPU GNN framework (\pname{}). 
\pname{} accelerates the feature aggregation with limited storage bandwidth by significantly improving hardware resource utilization, including GPU memory, CPU memory, and interconnect bandwidth, in multi-GPU systems. 
\pname{}  leverages a novel communication layer that allows GPU software caches to operate as a system-wide shared cache. 
This approach avoids the use of high-overhead, system-scope operations, aiming to maximize collective cache capacity and minimize redundant storage accesses while maintaining peak software cache bandwidth.

Furthermore, \pname{} enhances GPU software cache utilization through a hybrid eviction policy. 
This policy smartly evicts cold cache-lines by exploiting both the normalized reverse page-rank value of each node (static information) and the next reuse iteration (dynamic information). 
To gather dynamic information, \pname{} pre-executes a configurable number of graph sampling stages, thereby tracking the nodes sampled in subsequent iterations.

Finally, we introduce the novel Preemptive Victim-buffer Prefetcher (PVP) to efficiently prefetch previously evicted node feature data. 
Within \pname{}, when node feature data is evicted from the software cache, the cache line, along with its dynamic information, is moved to a victim buffer in CPU memory. 
It is then prefetched back to GPU memory as needed. Given that PVP operates without requiring GPU resources and can be executed asynchronously, it facilitates prefetching during the model training stage, a period when GPU PCIe ingress bandwidth is significantly underutilized.

When compared to the previous distributed training approaches, the optimized caching scheme in \pname{} enables each GPU to find more of its required graph data in its GPU memory and/or local CPU memory (victim cache), which helps improve the speed, efficiency, and scalability of multi-GPU GNN training pipelines.
Overall, we make the following key contributions:
\begin{itemize}
    \item We introduce an efficient large-scale storage-based multi-GPU GNN training framework without the requirement of higher storage bandwidth. 
    \item We design a novel communication layer that orchestrates the GPU's independent software caches into a shared cache that further lowers the bandwidth pressure on storage, without using the expensive (low-throughput) system-scope operations.  
    \item We present a hybrid cache eviction policy that leverages both static graph information and dynamic node access patterns to significantly improve cache hit ratios, thus reducing the pressure on the storage.
    \item We propose the PVP, which efficiently moves the evicted node feature data likely to be reused into CPU memory, enabling effective prefetching that further lowers storage contention.  
\end{itemize}

%% file: background.tex
\section{BACKGROUND}
\subsection{GNNs and GNN Training Pipeline}

Inspired by the success of convolutional neural networks (CNNs)~\cite{CNN0}, people started to apply similar filters to graphs \cite{lecunGCN, hamilton2017inductive, ying2019pinsage, GCNPierre, kipf2017semi, kipf2016variational, pmlr-v48-niepert16} and refer to such approaches as Graph Neural Networks (GNNs).
In GNNs, the graph adjacency matrix and node features are used as input during forward propagation.
For example, the forward propagation formula of a GCN layer is defined as $h^{out} = \sigma\left(A^{*}h^{in}W^{(l)}\right)$, where $h^{in}$ is the input node features, $h^{out}$ is the output node features, $W$ is the weights in this layer to be learned during the training process, and $A^{*}$ is the adjacency matrix normalized with the in degree and out degree of nodes.

\begin{figure}[!b]
\centerline{\includegraphics[width=0.8\linewidth]{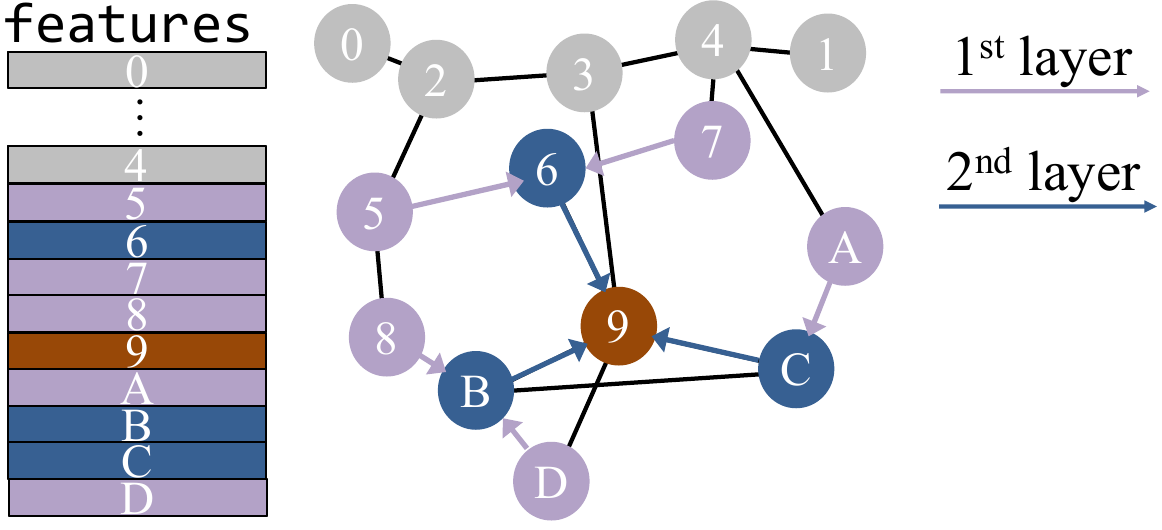}}
\caption{An example demonstrating the GraphSAGE neighbor sampling approach for output node 9. There are two layers in the model. On the left, the data layout of node attributes tensor, \texttt{feats}, of this graph is shown.}\label{fig:GrpahSAGE}
\end{figure}

GraphSAGE \cite{hamilton2017inductive} proposes neighbor sampling along with minibatch, greatly reducing the memory footprint.
A GraphSAGE model includes two to three layers, which can be mean, pooling, LSTM, etc.
Figure~\ref{fig:GrpahSAGE} shows an example of neighbor sampling on node 9.
Node indices are represented in hexadecimal.
Neighbors and 2-hop neighbors of node 9 are sampled, constituting the input of the second layer and input of the first layer.
%
%
%
Consequently, the sampled node features are scattered in the node feature tensor, as the illustration on the left shows. Graph Attention Network (GAT)~\cite{GAT} introduces learned self-attention masks into GraphSage layers to further enhance inference accuracy and will be the model used for evaluating the proposed optimizations in \pname.

Due to the large scale of graphs in many enterprise use cases, GNN model training usually adopts mini-batching. Each iteration in mini-batch training consists of three major stages: sampling, feature aggregation, and model training. 
In the sampling stage, the system samples the nodes and edges to create the subgraph for a mini-batch. Then, the feature data for the sampled nodes and edges are aggregated into a dense feature tensor in the feature aggregation stage. The sampled subgraphs with the feature data tensor are transferred to the GPU for the model training stage to produce gradients and update the model parameters.

\subsection{Distributed/Multi-GPU GNN Training}
\label{sec:bg_dist_training}

For large-scale graphs, GNN training is often executed in multi-GPU or distributed systems to address the memory capacity limitation and decrease the E2E training time by parallelizing the training process with multiple GPUs. 
In this case, the graph is partitioned across all nodes or GPUs, and the graph data is transferred over the network when other nodes or GPUs request the data during sampling or feature aggregation stages. To parallelize the GNN training, mini-batches are often distributed to compute nodes where each node has a full replicate of all the model weights and executes the forward propagation and backward propagation individually~\cite{distdgl} and then aggregates the model weights. 
Another approach is leveraging model parallelism where the model is split into several parts and different computing nodes compute each part~\cite{imagenet, model_parall1}.

The state-of-the-art distributed/multi-GPU GNN training encounters two major challenges. First, it incurs an extremely high cost for large-scale GNN training, as the cumulative memory capacity of GPUs or CPUs must surpass the total size of the graph dataset, which is often orders of magnitude higher than GPU memory capacity.  
Second, the partitioning of graphs across multiple nodes and GPUs introduces significant network communication overhead during the feature aggregation phase. To mitigate the communication costs, previous studies have employed the METIS-based graph partitioning algorithm~\cite{distdgl, wangFlexGraphFlexibleEfficient2021,dorylus}. 

However, graph partitioning algorithms, e.g., METIS, incur substantial preprocessing overhead. For example, it takes 457.55 seconds for DistDGL to partition the heterogeneous IGBH-medium~\cite{IGB} graph, which only involves 26.0M nodes and 249M edges, into four partitions on a node with 4$\times$ AMD ``Interlagos'' CPU. By contrast, under the same configurations, the random partition only costs 83.43 seconds. Additionally, the runtime of METIS-based algorithms scales exponentially with the size of the graph. Previous studies~\cite{linComprehensiveSurveyDistributed2022, caiDGCLEfficientCommunication2021} have proposed optimizations to reduce network communication overheads by overlapping communication with computation or smartly planning communication before actual transfer. However, the network communication overhead remains the main bottleneck, leading to grossly low GPU utilization.

Another approach to mitigate the communication overhead in the distributed GNN training is to leverage static GPU cache.

\subsection{Storage-based GNN Training}
In contrast to partitioning graphs across multiple nodes and GPUs, several studies~\cite{GIDS, Ginex, MariusGNN} have explored enabling GNN frameworks to fetch data directly from storage systems. Ginex~\cite{Ginex} and Marius GNN~\cite{MariusGNN} aim to hide the storage latency through in-memory caching and pipelining of storage accesses using CPU orchestrated approaches while GIDS~\cite{GIDS} enables GPU threads to access feature data directly, leveraging the massive parallelism of GPUs to mitigate storage latency. Unlike conventional distributed GNN training frameworks, storage-based GNN training obviates the need for additional resources or graph partitioning. However, none of these previous studies support GNN training in multi-GPU systems as the feature aggregation performance is dependent on the storage bandwidth, which may not be scalable with the number of GPUs or nodes in the system in practice.

\subsection{Scoped Memory Consistency Model}

Since Volta architecture~\cite{PTXFormal}, NVIDIA GPUs use the Scoped Memory Consistency Model. NVIDIA's Scoped Memory Consistency Model is weakly ordered and utilizes scoped synchronization primitives like threading block (\texttt{.cta}), at the device or current GPU (\texttt{.gpu}), and across the whole system (\texttt{.sys}) levels to facilitate interthread communication through memory, allowing threads within the same thread block to synchronize efficiently while not mandating data race freedom, resulting in a more intricate set of rules. 
Concurrent algorithms can exploit memory consistency operations using the GPU memory model to select if they want to quality with the \texttt{.acquire}, \texttt{.release} semantics or \texttt{.relaxed} or \texttt{.weak} in addition to the memory scope to improve their performance. This work exploits these memory consistency capabilities to achieve higher performance when accessing cache metadata.

%% file: motivation.tex
\section{MOTIVATION}

The state-of-the-art storage-based GPU GNN training systems face significant scalability challenges in multi-GPU settings due to limited storage bandwidth, particularly during the feature aggregation stage.

\begin{figure}[bt]
    \includegraphics[width=0.9\columnwidth]{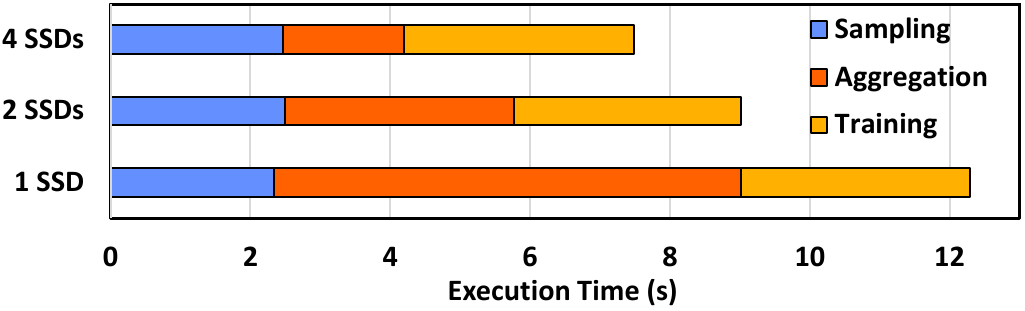}
    \caption{Breakdown of GNN training time for storage-based GNN training, illustrating the effect of varying the number of SSDs. The feature aggregation time with a single SSD connected is 3.8$\times$ higher than with four SSDs connected. One GPU is used in this measurement.}
    \label{fig:ssd_scale}
\end{figure}

Figure~\ref{fig:ssd_scale} illustrates the disparity in feature aggregation times when employing 1, 2, and 4 SSDs for storage-based GNN training. Notably, the time required for feature aggregation with a single SSD is nearly four times longer than when four SSDs are used. As more GPUs are used, the pressure on the SSD will further increase and the aggregation stage for each GPU will take longer, negating the benefit of using more GPUs. This storage bandwidth bottleneck significantly hinders scalability as the system transitions from 1 to 4 GPUs with a fixed number of SSDs, emphasizing the need for a more efficient approach to scale the number of GPUs that can perform the feature aggregation stage simultaneously with a fixed number of GPUs.

Previous studies~\cite{GIDS, bam} address this problem 
by connecting multiple SSDs to a single GPU, thereby increasing the collective storage bandwidth available. However, this approach is not a practical approach in multi-GPU systems due to two main challenges. Firstly, achieving the necessary 
ingress bandwidth for all GPUs requires a linear increase in the number of SSDs as a function of the number of GPUs. For instance, fully matching the intake bandwidth of one A100 GPU with the high-performance Intel Optane SSDs necessitates five SSDs per GPU, scaling to 40 SSDs for an eight-GPU system. This requirement becomes even more daunting with less performant SSDs like those from Samsung 980pro, potentially exceeding 60 SSDs. Secondly, accommodating changes in SSD configurations necessitates either duplicating or reformatting data across all SSDs, a process that introduces considerable overhead when a large number of SSDs are involved, especially given the large scale of graph data involved in GNN training.

\begin{figure}[bt]
    
    \includegraphics[width=\columnwidth]{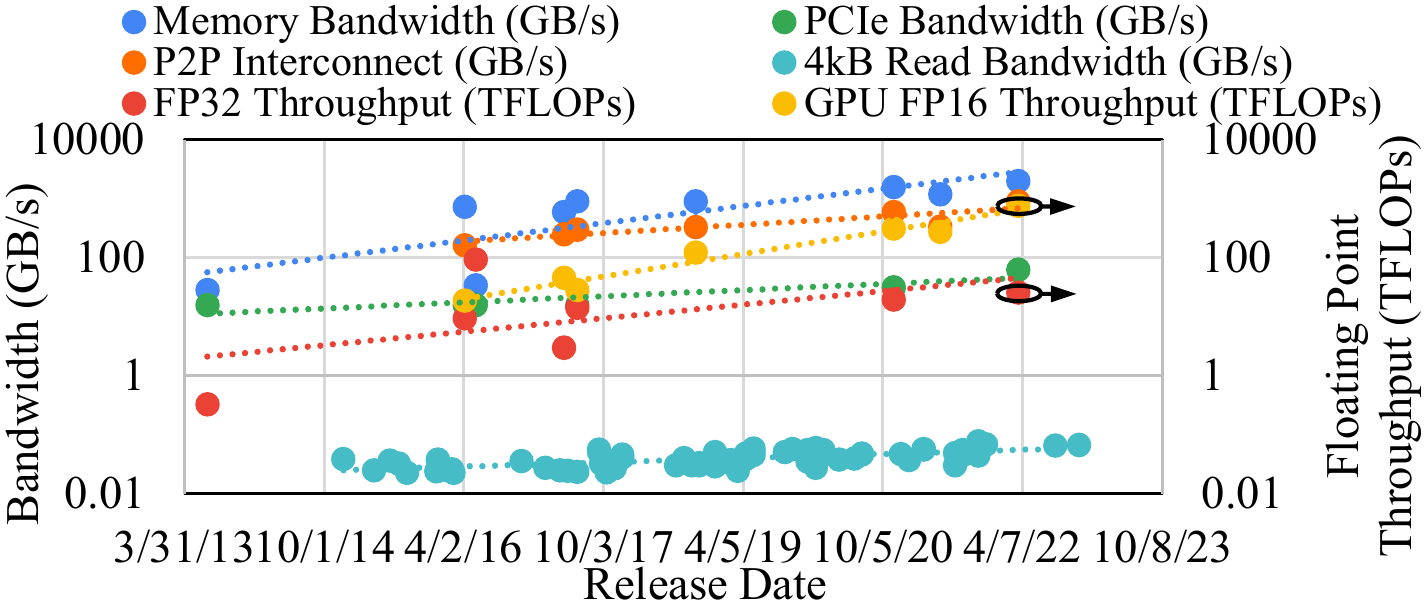}
    \caption{Trend of recent GPUs for deep learning and consumer-grade SSDs. Compared with other select metrics, the bandwidth of SSDs has the lowest growth rate. We collect the inter-device (P2P) bandwidth, PCIe bandwidth, memory bandwidth, and floating-point throughput of Nvidia 100-level GPUs since Kepler (K100) and Google TPUs~\cite{epochParameterComputeData,wikipediaTensorProcessingUnit2024,jouppiInDatacenterPerformanceAnalysis2017,techpowerupGPUSpecsDatabase2024}. We sampled the 4kB read bandwidth of recent consumer-grade SSDs with 1TB capacity from UserBenchmark~\cite{userbenchmarkSSDUserBenchmarks1072}.
     }
    \label{fig:trend}
\end{figure}

The challenge of limited storage bandwidth is anticipated to intensify based on the growth trend of interconnect and hardware resources over recent decades. Figure~\ref{fig:trend} compares the growth rates of SSD read bandwidth, PCIe bandwidth, GPU memory bandwidth, and GPU compute throughput~\cite{userbenchmarkSSDUserBenchmarks1072, epochParameterComputeData,wikipediaTensorProcessingUnit2024,jouppiInDatacenterPerformanceAnalysis2017,techpowerupGPUSpecsDatabase2024}. The slowest growth rate of storage bandwidth starkly indicates that the challenge of limited storage bandwidth in multi-GPU settings will become increasingly critical for storage-based multi-GPU GNN training systems.

To this end, we develop the \pname{} that efficiently utilizes available resources to scale the feature aggregation stage with limited storage bandwidth.

%% file: design.tex
\section{\pname{} SYSTEM DESIGN}

\begin{figure}[htbp]
    
    \includegraphics[width=\columnwidth]{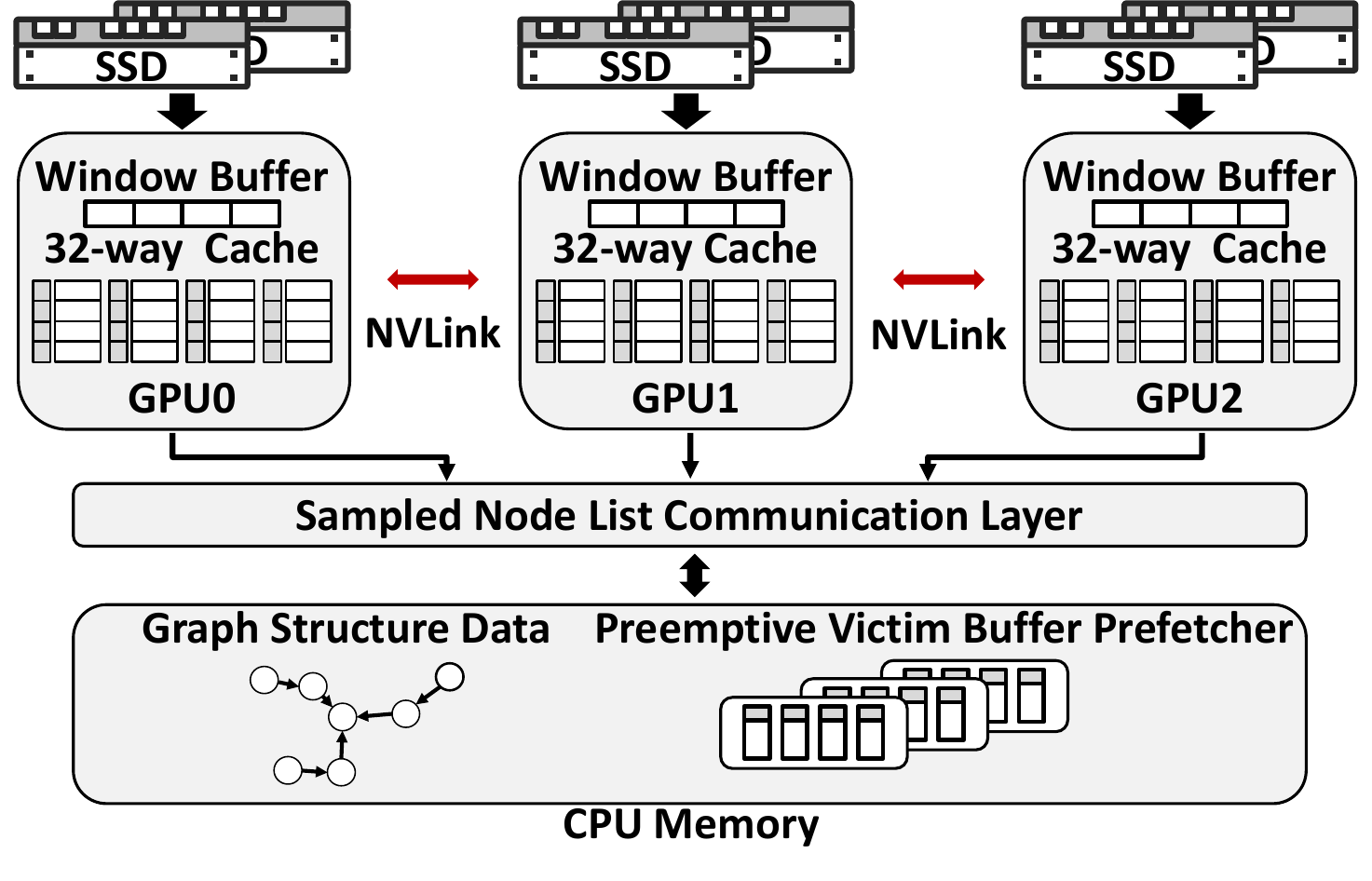}
    \caption{Illustration of the System Design of \pname{}.}
    \label{fig:overview}
\end{figure}

The overall system design of \pname{} is illustrated in Figure~\ref{fig:overview}. As \pname{} is a storage-based GNN training framework, feature data is stored in SSDs and directly fetched by GPU threads. To reduce the pressure on the storage, \pname{} employs a 32-way set associative GPU software cache to temporarily store the feature data of recently accessed nodes. \pname{} leverages the window buffering technique from GIDS~\cite{GIDS} to store the list of sampled nodes to be used in the next iterations in the window buffer, which provides node access pattern information to the cache and allows the node features to be used in these future iterations to be preserved if they are evicted from the cache. To orchestrate the individual GPU software caches into 
a shared cache, \pname{} has a communication layer for GPUs to receive their needs from other GPUs' software caches. 
Finally, the graph structure data is pinned in the CPU memory (system memory) to enable GPU threads to directly fetch graph data with Unified Virtual Addressing (UVA) during graph sampling while Preemptive Victim Buffer Prefetcher (PVP) stores reusable evicted cache-lines from GPU and transfers them back to GPU when they are needed.

\subsection{Communication Layer 
} 
Efficiently managing GPU memory across multiple GPUs is critical for increasing the effective bandwidth of feature aggregation beyond limited storage bandwidth in a storage-based GNN training framework.
Similar to previous studies~\cite{GIDS, bam}, \pname{} utilizes GPU memory as a software cache to capture the spatial and temporal locality of node access patterns to minimize storage accesses.

However, due to the randomness of the sampling process, cache capacity becomes a critical factor in achieving a high cache hit ratio.
In a multi-GPU environment, achieving this involves orchestrating the GPUs' software caches into a shared cache.  The node features are shared and assigned to the GPUs so that the node features can reside in more than one GPU's software cache. 
The benefits of the shared cache system are particularly substantial in systems equipped with NVLink, as NVLink offers significantly higher bandwidth compared to traditional storage and PCIe bandwidth.

A naive implementation of the shared cache would enable each GPU to directly access and manage each other's software cache. 
However, this approach faces two major challenges. 
First, any changes to the data in the shared cache must be visible to all GPUs, requiring system-scope cache management operations to ensure memory consistency.
These cache management operations require system-scope memory operations, whose latency is significantly higher and throughput is significantly lower than their device-scope counterparts, limiting the effective bandwidth of the cache.

\begin{figure}[h]
    
    \includegraphics[width=\columnwidth]{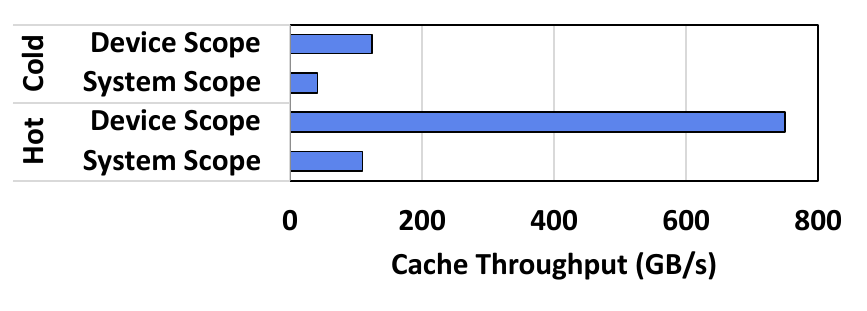}
    \vspace{-6ex}
    \caption{Comparison of GPU software cache effective bandwidth using two scopes: device and system for cache management operations. }
    \label{fig:scope_bench}
\end{figure}

Figure~\ref{fig:scope_bench} illustrates the impact of cache management operation scope on the software cache's effective bandwidth.
As shown, the effective bandwidth of the cache for cache hit (Hot) is around 750 GBps with the device-scope memory operations and drops to 110 GBps with the system-scope memory operations.
Similarly, for cache misses (Cold), the cache achieves an effective bandwidth of 124 GBps for device-scope memory operations and only 42 GBps for system-scope memory operations. 
These results demonstrate how the implementation of shared cache design with system-scope operations substantially limits the cache access bandwidth.

The second issue with the naive approach to implementing a shared cache arises from the overhead associated with the Python Global Interpreter Lock (GIL). In order to facilitate a basic shared cache, GPU threads must be able to directly access caches on other GPUs to retrieve cache lines. This necessitates that GPU threads share the same address space, implying that multi-GPU training should utilize multi-threading, with each thread dedicated to managing a single GPU. However, both DGL and PyG advocate for the use of Distributed Data Parallel (DDP) over Data Parallel. DDP leverages multi-processing parallelism, which is preferred due to the substantial overhead incurred by multi-threading parallelism as a result of the GIL~\cite{DDP}.



\begin{figure}[t]
    
    \includegraphics[width=0.9\columnwidth]{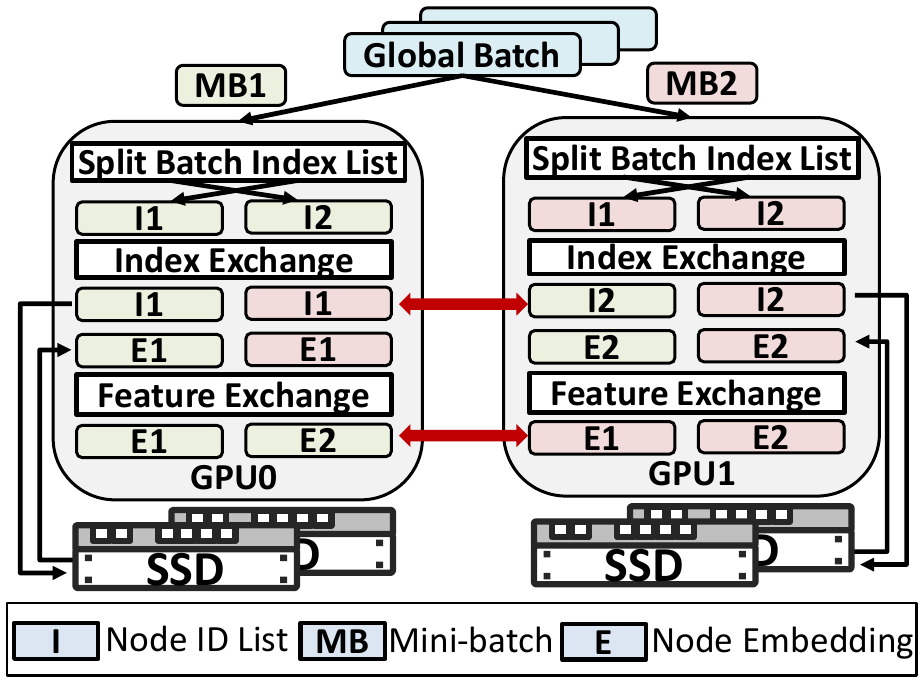}
    \caption{The Communication Layer process with 2 GPUs.}
    \label{fig:system_layout}
\end{figure}

To address these problems, we propose a novel communication layer that orchestrates GPUs' software caches into a system-wide shared software cache.
Figure~\ref{fig:system_layout} illustrates the stages within this layer. 
Initially, each GPU generates a list of sampled nodes for its mini-batch through the sampling process and then splits the list into sub-lists using a hash function, which determines the cache-lines assigned to each GPU software cache. 
Although our implementation utilizes a straightforward striding function, more complex hash functions, such as those derived from graph partitioning algorithms, could further reduce communication overhead. 

Each GPU then communicates its sub-lists to its assigned GPUs. Since each sub-list consists of only node IDs the overhead of this communication is negligible. 
Each GPU then independently performs the feature aggregation process for the sampled nodes requested by all GPUs. This pre-distribution of accesses to the GPU software caches ensures only the local GPU threads are accessing the software cache, eliminating the need for system-scope memory operations. After feature data is fetched, they are transferred to the requesting GPU which assembles the original mini-batch corresponding to the sampled nodes, ensuring the model training stage remains unaffected by this optimized communication layer.

To facilitate this system, a global barrier synchronization 
is introduced after the node ID lists are divided, ensuring all GPUs progress in lockstep. While such synchronization might typically introduce significant overhead in multi-GPU setups, in our context, it proves minimally disruptive. This is because GPUs are already synchronized post-training during gradient accumulation, which minimizes the latency of the additional barrier synchronization. 
Furthermore, the communication layer abstracts the storage layout, allowing for flexible storage configuration tailored to specific hardware setups, thereby enhancing the system's adaptability and efficiency.

\subsection{Hybrid Eviction Policy}
\label{design:hybrid_eviction_policy}
The cache hit ratio during the feature aggregation process for large-scale GNN training often remains low, due to the vast size of graph datasets and the inherent randomness of the sampling process. 
Thus, it is crucial to incorporate application-specific information to guide the software cache toward more efficient exploitation of data locality.

The cache can leverage two types of information to efficiently exploit locality. The first type is static information about the graph structure. Previous studies~\cite{DataTiering} demonstrated that the static information of the graph, such as out-degree or weighted reverse page rank score, can effectively distinguish frequently accessed nodes from less frequently accessed nodes. Thus, the software cache can leverage this information to minimize evicting frequently used cache lines.

The second type is dynamic information. Previous work,
GIDS~\cite{GIDS}, leverages the timing flexibility of the graph sampling process to gather 
the list of nodes sampled in upcoming iterations and store them in the window buffer. 
To avoid adding latency to the cache eviction operations, \pname{} scans the sampled nodes in the window buffer to determine the next reuse iteration for the cache-lines that currently reside in the cache before the feature aggregation stage.

Although dynamic information provides accurate critical information for the node access pattern, the cost of scanning the next reuse time for all cache-lines in the cache can introduce significant overhead, especially for large window buffer sizes (number of iterations to foresee). 
To mitigate the impact of this overhead, \pname{}'s software cache updates the dynamic information for a configurable number of iterations, rather than for every iteration.  The frequency of such updates is determined based on the cache-size, batch size, and window buffer size.

Both static and dynamic information offer insights into the unique characteristics of node access patterns. Thus, \pname{} improves the cache efficiency of GPU software by implementing a hybrid eviction policy that leverages both types of information.
The hybrid eviction policy classifies each eviction candidate into four priority levels based on their dynamic information. If there is more than one eviction candidate at the same priority level, the eviction policy selects the one with the lowest static priority value for eviction. The eviction policy first searches for candidates that are in the lowest priority level and goes to the next level if there are no candidates in the current level.  

The lowest priority is assigned to nodes with no anticipated reuse within the window buffer, suggesting that their feature data will not be required in the foreseeable future. Thus, the probability of reusing these cache lines is low. 

The next priority level is the nodes whose dynamic information value falls below a specific threshold. This threshold is configurable, but by default, it is defined as 1/8 of the window buffer size. This priority level indicates that the candidates are expected to be reused within the window buffer but not in the immediate future, as suggested by the threshold value.

Candidates at the third priority level consist of recently inserted nodes. Since these nodes have just been added to the cache, they lack dynamic information. Given their recent insertion, there's a potential for these nodes to be reused in subsequent iterations, granting them a higher priority over the first two groups of candidates.

The highest priority is assigned to nodes whose reuse value exceeds the specified threshold. These candidates are expected to be reused in the near future, making it crucial to retain them in the cache. 

This hierarchical approach ensures efficient cache management, prioritizing the retention of nodes likely needed in the imminent iterations while facilitating the eviction of those less likely to be reused. Moreover, static information helps the eviction policy to minimize and avoid evicting hot nodes even if they fall into lower priority levels. The flexibility of the hybrid policy allows adjustment of thresholds and priority levels, tailoring cache behavior to the specific demands of the GNN training process.

\subsection{Preemptive Victim-Buffer Prefetcher}
\label{sec:pvp}

Traditional GNN feature data prefetchers, which are primarily based on the CPU, attempt to preload feature data for upcoming iterations while the current iteration is being processed by the GPU. However, as evidenced by previous studies~\cite{pytorch-direct, GIDS}, these prefetching methods fall short of matching the GPU's model training throughput, leading to 
little performance improvement. 
Moreover, attempts at GPU-based prefetching can lead to resource contention due to the significant GPU resources, e.g., GPU memory, required to exploit the GPU's massive parallelism to hide the storage latency as GPU resource is already utilized for model training. This challenge has rendered previous prefetching approaches ineffective within storage-based GNN training frameworks.

Thus, \pname{} enhances the efficiency of the feature aggregation phase with the introduction of the PVP, aimed at optimizing interconnect bandwidth utilization across the GNN training pipeline. The PVP is specifically designed to improve bandwidth usage during periods when PCIe bandwidth is underutilized, such as during the model training stage without increasing GPU resource contention. It achieves this by preemptively managing the eviction and prefetching of cache lines, guided by dynamic information.

\begin{figure}[t]
    \centering
    \includegraphics[width=\columnwidth]{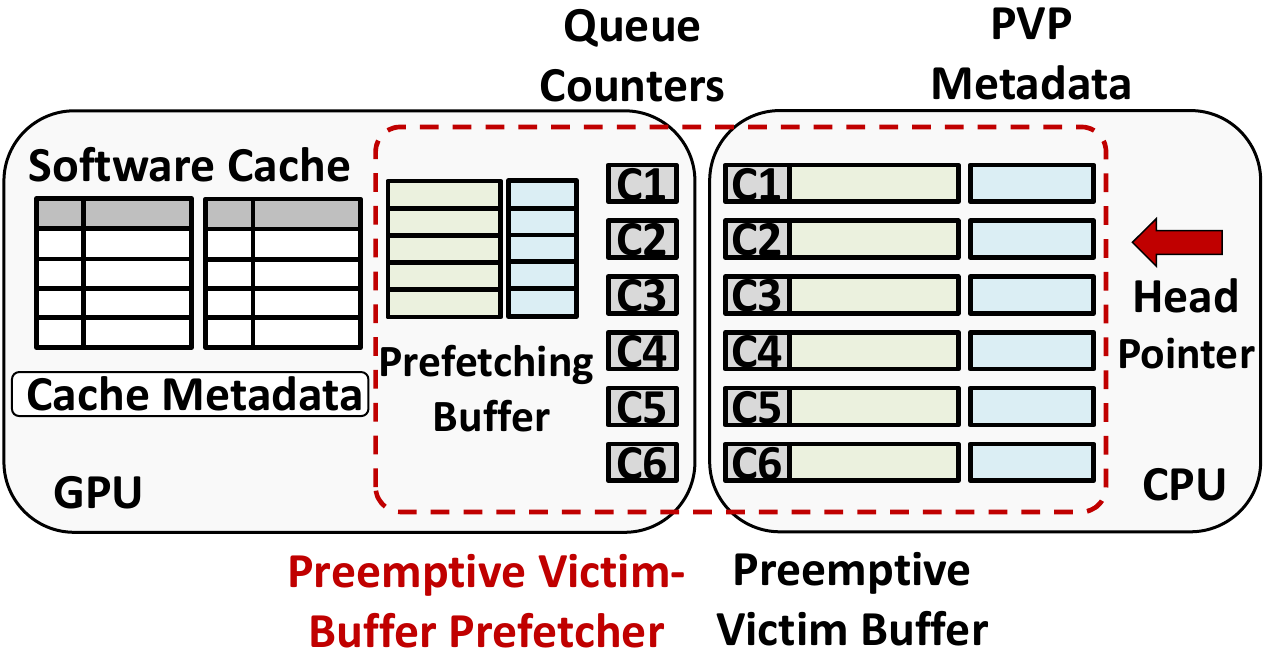}
    \vspace{-4ex}
    \caption{Preemptive Victim-Buffer Prefetcher (PVP). 
    }
    \label{fig:pvp_architecture}
\end{figure}

Figure~\ref{fig:pvp_architecture} illustrates the architecture of the PVP. On the CPU side, the architecture includes data and metadata victim buffers tasked with managing evicted cache-line data and their associated metadata, respectively. This setup ensures efficient handling of data that is no longer retained within the GPU's cache.
On the GPU side, the PVP is equipped with counters to track the number of victim cache-lines present in the buffer. GPU threads use atomic operations to increment these counters each time an evicted cache-line is enqueued into a victim buffer. Additionally, a prefetching buffer is designated for the temporary storage of prefetched data. 
This setup and the PVP Metadata on the CPU side enable CPU threads to directly transfer data to the appropriate entries of the prefetching buffer on the GPU side through a CUDA API call (\texttt{cudaMemcpyAsync}), streamlining the prefetching process without 
access to the GPU's software cache. Moreover, the CPU to GPU transfer is done at the training stage of the pipeline, when the PCIe ingress bandwidth is not utilized and thus does not cause extra resource contention.

\subsubsection{Eviction Process with PVP} 

\begin{figure}[t]
    \centering
    \includegraphics[width=\columnwidth]{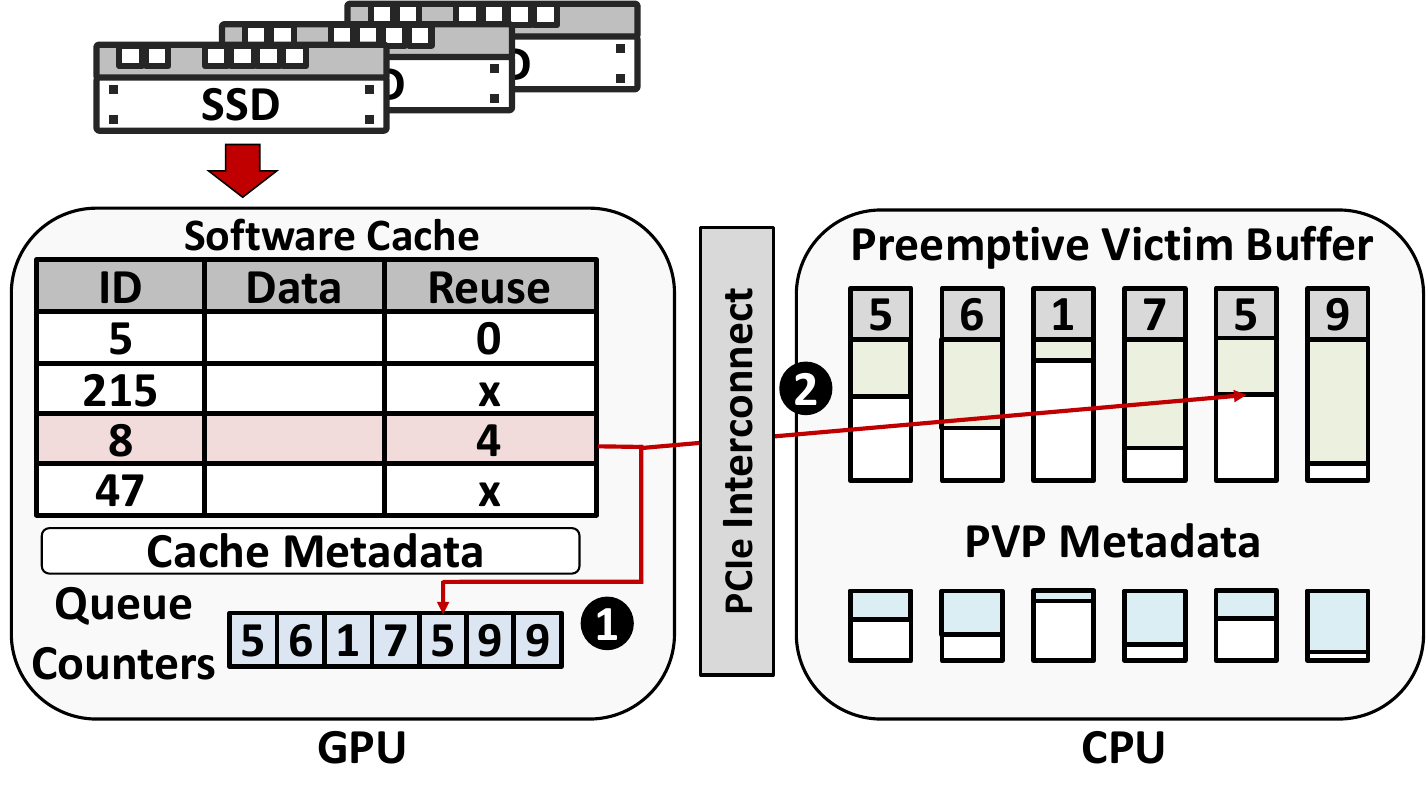}
    \caption{Example of the eviction process with PVP.
    }
    \label{fig:pvp_evict}
\end{figure}

The \pname{} PVP introduces an additional eviction stage during the feature aggregation stage, leveraging the next reuse timestamp for each node's feature data in the \pname{} GPU software cache. Figure~\ref{fig:pvp_evict} illustrates this eviction stage with PVP.

Upon 
evicting a node's feature data from the GPU software cache, the leader thread first checks the next reuse iteration. If the cache-line has the information for the next reuse iteration, a GPU thread calls an atomic operation to increment the corresponding queue counter (1), which tracks the number of evicted node feature data currently stored in the victim buffer. 
This incremented counter value identifies the appropriate index for storing the data within the corresponding victim buffer. In scenarios where the counter value surpasses the buffer's capacity or if no reuse iteration is marked, the cache-line data is simply discarded. 
Otherwise, the obtained counter value is shared among the threads in the same warp, guiding the collective transfer of the cache line into the prefetcher buffer in the CPU memory (2). For instance, with a next reuse timestamp of 4 and a queue counter at 5, the evicted cache line is placed into the 5th position of the fourth victim buffer.

Importantly, the additional eviction stage does not increase resource contention. GPU PCIe egress bandwidth is not utilized during the feature aggregation stage for the GPU-oriented feature aggregation process. Thus, there is no contention on the PCIe egress bandwidth. 
Moreover, since the number of concurrently evicting cache lines is proportional to the number of accesses to the storage, the latency of evicting cache-line to the CPU victim buffer is hidden when there are a sufficient number of storage accesses.

\subsubsection{Eviction Policy with PVP} 

In Section~\ref{design:hybrid_eviction_policy}, we introduce a hybrid eviction policy designed to improve the GPU software cache hit ratio by utilizing both static and dynamic information. However, when PVP is activated, a slight modification to this policy is necessary to fully harness its performance benefits.

Under the original hybrid eviction policy, cache-lines not expected to be reused within the window buffer's scope are prioritized for eviction. These cache-lines, lacking tags indicating their next reuse iteration, are directly discarded. 
This approach can lead to missing potential reuse opportunities outside the window buffer scope. 
To address this, \pname{} swaps the two lowest priority levels in the hybrid eviction policy when PVP is enabled.  This modification prioritizes the cache lines tagged with the next reuse iteration but below the threshold for the eviction so that they can be evicted into the victim buffers. This modified policy allows more of the cache-lines, which are not reused within the window buffer's scope, to remain in the software cache.
This strategy enhances both PVP efficiency and overall GPU cache utilization by ensuring that cache-lines with foreseeable reuse but below the threshold are recycled through the PVP path.

\subsubsection{PVP's Metadata} 
Once victim cache-lines are prefetched and temporarily stored in the prefetching buffer, they must be copied back to the mini-batch. However, since the mini-batch is not sorted by node ID, locating the appropriate position for the prefetched data within the mini-batch would necessitate a search operation, potentially introducing significant overhead. To mitigate this, the system manages metadata including the batch ID, GPU ID, and next reuse iteration within the software cache. This metadata is evicted to the victim buffer alongside the cache-line when it is moved to the victim buffer. Consequently, when GPU threads retrieve data from the prefetching buffer, they efficiently identify the corresponding batch index for placing the feature data into the appropriate entry of the mini-batch data buffer. 

To minimize memory usage and reduce the number of operations required during the dynamic information update stage, PVP adopts a specific metadata representation. The first 2 bytes indicate the next reuse iteration, setting a limit on the window buffer size to no more than 65,536. The next byte identifies the GPU ID that requested the sampled node. The final 5 bytes denote the index within the mini-batch. Since the initial 2 bytes represent the reuse time, atomicMin operations are utilized during the dynamic information update stage to record the next reuse iteration efficiently.

%% file: Evaluation.tex
\section{EVALUATION}
\subsection{Experimental Setup}

\textbf{Environment.} We compare \pname{} and the state-of-the-art baseline storage-based GNN framework using the system described in Table~\ref{tab:config}. This system has 2$\times$ NVIDIA A100-40GB GPUs and the GPUs are connected with 3 NVLink bridges.
%
For the baseline distributed GNN framework, we evaluated on a two nodes system where each node is equipped with 2$\times$ AMD ``Milan'' EPYC 7763 CPU, and 2TB memory. Each node has 2$\times$ A100 PCIe 40GB, with PCIe Gen 4 and NVLink. 

{\renewcommand{\arraystretch}{1.2}
\begin{table}[h]
\centering
\scriptsize
    \caption{\small {Configuration used to evaluate \pname{}.}}
    \vspace{-2ex}
\begin{tabular}{|p{0.8in}|p{2.0in}|}
    \hline
    {\textbf{Configuration}}& {\textbf{Specification}} \\
    \hline
    \hline
	{CPU}                  & {AMD EPYC 7702 64-Core Processor} \\
    \hline
	{Memory}               & {1TB DDR4} \\
    \hline
	\multirow{1}{*}{GPU} &  NVIDIA A100 HBM2 40GB \\ 
    \hline
	{NVLink}               & {3 bridges, 270 GBps peak Bandwidth} \\
    \hline
    \multirow{2}{*}{S/W}   & Ubuntu 20.04 LTS, NVIDIA Driver 470.103 \\
                           & CUDA 12.1, DGL 2.0.0, Pytorch 2.0.1 \\
    \hline
    \multirow{1}{*}{SSDs}     &   Intel Optane SSDs, PCIe Gen 4 Interconnect \\
    \hline
\end{tabular}
\label{tab:config}
\end{table}
}

\textbf{Datasets.} To assess the performance of \pname{} on large-scale graph datasets, we conducted experiments using five real-world datasets: IGB-Full~\cite{IGB}, IGBH-Full~\cite{IGB}, IGB-Medium~\cite{IGB}, ogbn-papers100M~\cite{ogbn_paper}, and MAG240M~\cite{MAG} as shown in Table~\ref{tab:dataset}. As ogbn-papers100M and MAG240M datasets are relatively small,  we only tested with a 4 GB GPU software cache size.  We used the IGB-Full dataset by default in the evaluations if not explicitly mentioned.

{\renewcommand{\arraystretch}{1.2}
\begin{table}[h]
\centering
\scriptsize
    \caption{\small Real-world dataset used for evaluating \pname{}. 
}

\begin{tabular}{|p{0.6in}|p{0.54in}|>{\raggedleft\arraybackslash}p{0.35in}|>{\raggedleft\arraybackslash}p{0.35in}|>{\raggedleft\arraybackslash}p{0.52in}|}
    \hline
    \textbf{Dataset} & \textbf{Graph Type} & \textbf{\# Nodes}& \textbf{\# Edges}& \textbf{Feature Size}  \\
    \hline
    \hline
	papers100M & Homogeneous & 111M  & 1.62B & 128 \\
    \hline
	IGB-Full & Homogeneous &  269M & 4.00B &  1024 \\
    \hline
        IGB-Medium & Homogeneous &  10.4M & 120B &  1024 \\
    \hline
	MAG240M & Heterogeneous & 244M & 1.73B &  768 \\
    \hline
	IGBH-Full & Heterogeneous & 547M & 5.81B & 1024  \\
    \hline
\end{tabular}
\label{tab:dataset}
\end{table}
}


\textbf{Model:} We assessed \pname{}'s performance with neighborhood sampling~\cite{Graphsage} with fanout values (5,2,2,2) and 4-layer GAT~\cite{GAT} model whose hidden channel dimension is 512. Finally, we set the batch size as 2048 for all evaluations. 

\textbf{\pname{} Configuration:} In the default configuration, we allocated 16 GB of GPU device memory for each GPU software cache.
For the hybrid eviction and dynamic eviction policies, the default configuration is 256 iterations for the window buffer size with 4 iterations for the window buffer update period. 
For PVP, each Victim buffer in the CPU memory can capture up to 16K cache-lines.

\textbf{Measuring Execution Time:} When working with large graph datasets, the training process can be excessively long. Therefore, we conducted the evaluations by measuring the execution time for a sum of 100 iterations after a warm-up stage of 1,000 iterations. 

\textbf{Baseline: } To compare \pname{} with storage-based GNN frameworks, we extended GIDS to work in multi-GPU systems (M-GIDS), which does not incorporate any of the \pname{} techniques. To reduce cache metadata for the GPU software cache, we integrated a 32-way set associative cache with a Round-Robin eviction policy into M-GIDS. For distributed GNN training, we compared against Dist-DGL~\cite{distdgl}.

\subsection{Impact of the Communication Layer}
In this section, we investigate the effects of the Communication Layer on the GPU software cache hit ratios and feature aggregation times. Our evaluation compares \pname{} with the GIDS based multi-GPU GNN training framework (M-GIDS), and both systems are configured with two SSDs (one SSD per GPU). Performance metrics were gathered using two distinct eviction policies: the Round-Robin (RR) eviction policy and the hybrid (H) eviction policy.

For both \pname{} and M-GIDS, the GPU software caches are set at sizes of 4 GB, 8 GB, and 16 GB. The hybrid eviction policy was evaluated with a window buffer with a size of 256, and the Preemptive Victim-buffer Prefetcher was disabled for this assessment. Additionally, for \pname{}, the feature aggregation time includes communication time created by the communication layer.

\begin{figure}[tb]
    \vspace{1ex}
    \centering
    \includegraphics[width=0.9\columnwidth]{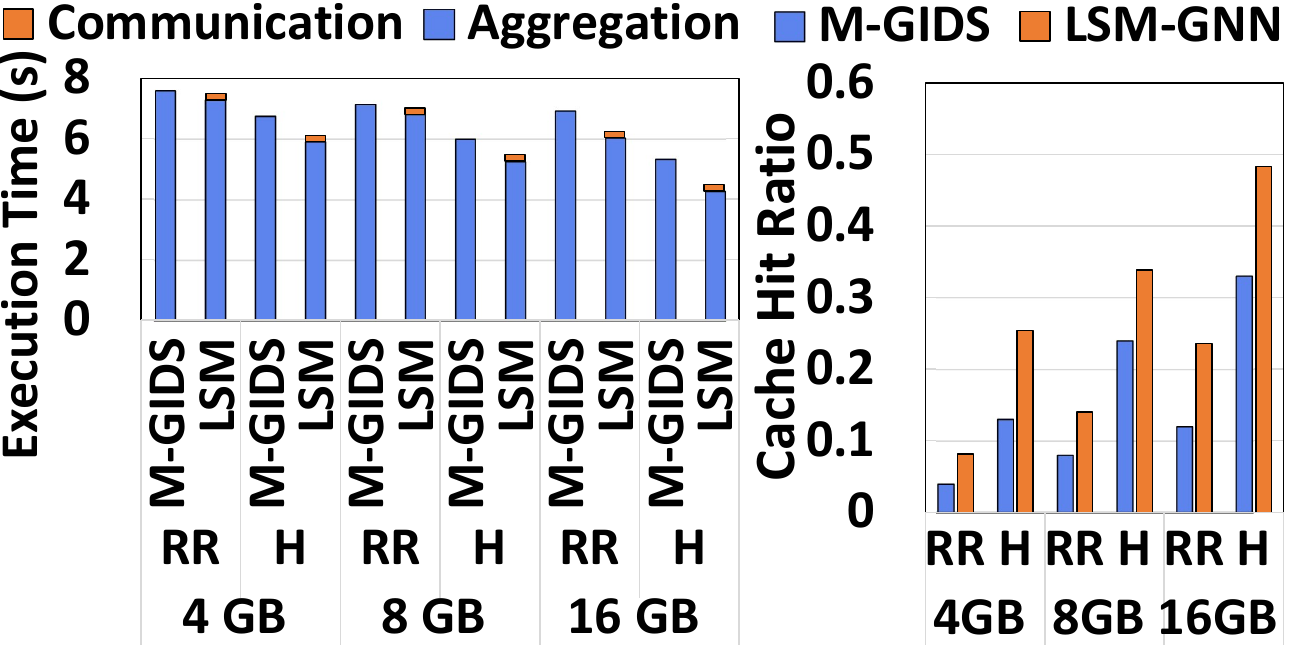}
    \caption{The impact of higher cache capacity achieved by the communication layer. \pname{} achieves up to 1.18$\times$ speed up for the feature aggregation compared to the baseline framework with independent GPU software caches mainly because of better cache hit ratios. }
    \label{fig:eval_com_layer}
\end{figure}

Figure~\ref{fig:eval_com_layer} illustrates the cache hit ratios and feature aggregation times for M-GIDS and \pname{} with the Communication Layer. As shown, the Communication Layer allows \pname{} to achieve cache hit ratios, and thus time, comparable to those of the baseline configuration with a cache size that is twice as large. This is achieved by enabling the GPU software caches to function collectively as a system cache. 

The feature aggregation time significantly decreases with larger cache sizes and with the hybrid eviction policy. When the cache size is set to 4 GB using the Round-Robin eviction policy, the feature aggregation times are similar. However, with a 16 GB cache under the hybrid eviction policy, the time is reduced by 1.18$\times$. This is because the feature aggregation time is proportional to the cache miss rate, and a larger cache with a more efficient eviction policy enables the cache to exploit spatial and temporal locality, resulting a higher differences in cache miss rate.

Furthermore, the communication time remains constant, irrespective of the cache capacity and hit ratio, accounting for only 0.21 seconds, which is less than 5\% of the feature aggregation time in the worst-case scenario. Nevertheless, for the communication layer to contribute to performance enhancement, the feature aggregation time must decrease by more than 5\% due to increased cache capacity.


\subsection{Impact of the Cache Eviction Policy}

This section explores the impact of different eviction strategies on cache performance. We evaluated the cache hit ratio and feature aggregation time using four distinct eviction policies: 1) Round-Robin, 2) Static Information-Based, 3) Dynamic Information-Based, and 4) Hybrid Information-Based eviction policies. In this evaluation, the reverse PageRank value represents static information, while the next reuse iteration serves as dynamic information. The static information-based eviction policy prioritizes evicting the cache line with the lowest static value, whereas the dynamic information-based eviction policy targets the cache line that has the longest time until the next reuse (a cache line with no expected reuse has the highest eviction priority). We conducted experiments with three different GPU sizes: 4 GB, 8 GB, and 16 GB, and two SSD configurations: 2 SSDs and 4 SSDs in the system. For all tests, we standardized the GPU software cache configuration to a 32-way set associative cache.

\begin{figure*}[htb]
    \centering
    \includegraphics[width=0.9\textwidth]{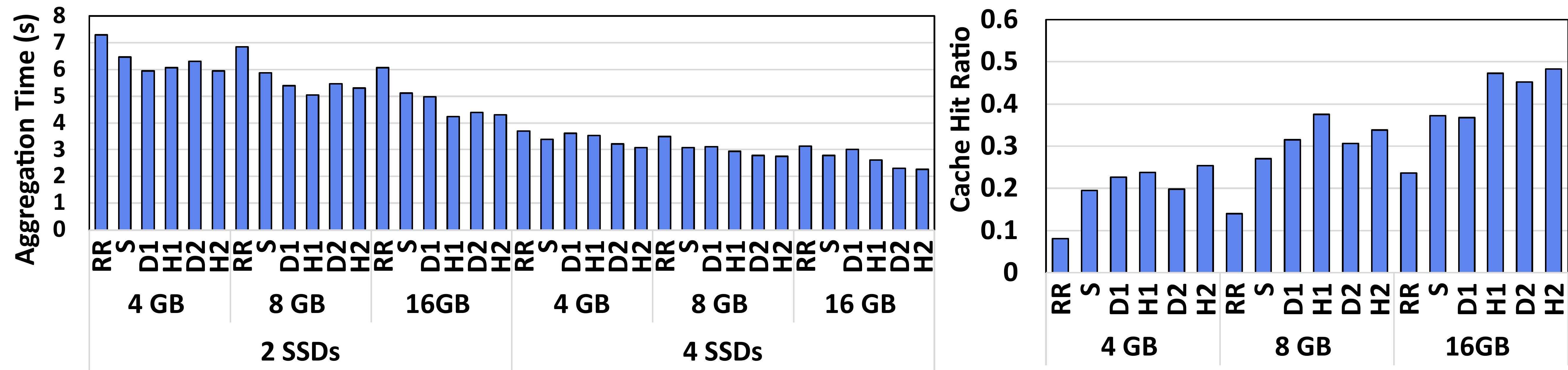}
    \caption{Impact of the cache eviction policies on \pname{} software cache. RR, S, D, and H stand for Round-Robind, Static information based, Dynamic information based, and Hybrid eviction policies, respectively. D1 and H1 have a smaller window buffer size (64) while D2 and H2 are evaluated with a window buffer size of 256.
    The hybrid eviction policy can increase the cache hit ratio from 0.23 to 0.48 when the cache size is 16 GB, resulting in 1.41$\times$ speed up for the feature aggregation process.}
    \label{fig:eval_cachebench}
\end{figure*}

Figure~\ref{fig:eval_cachebench} illustrates the hit ratios and feature aggregation times across these eviction policies. In all configurations, the hybrid eviction policy yields the highest cache hit ratio, while the Round-Robin policy results in the lowest cache hit ratio. The hybrid eviction policy significantly improves the cache hit ratio compared to the Round-Robin eviction policy, increasing it from 0.23 to 0.48 when the cache size is 16 GB and achieves a 1.41$\times$ speedup with 2 SSDs and a 1.38$\times$ speedup with 4 SSDs.

The performance gain from the hybrid cache eviction policy is greater in the case with 2 SSDs than with 4 SSDs. This is because the feature aggregation process is bounded by storage bandwidth. Thus, contention for storage accesses is more detrimental to the aggregation process when the storage bandwidth is lower. Thus, the performance gain will be higher in a system where a single SSD is connected to two GPUs or SSDs with lower read bandwidth. Furthermore, the performance gain of the hybrid eviction policy compared to the Round-Robin eviction policy is greater when the cache capacity is larger. For example, with 2 SSDs in the system, the feature aggregation process speed ups are 1.22$\times$, 1.29$\times$, and 1.41$\times$ for GPU software caches of 4 GB, 8 GB, and 16 GB, respectively. This is because a higher cache capacity enables the cache to capture more node access locality, increasing the advantage over the Round-Robin eviction policy.

The hybrid eviction policy outperforms both the static and dynamic information-based eviction policies in any setting. It specifically surpasses the static information-based policy as cache size increases. The static eviction policy achieves cache hit ratios of 0.19, 0.27, and 0.37, while the hybrid eviction policy attains 0.25, 0.33, and 0.48 for cache sizes of 4 GB, 8 GB, and 16 GB, respectively. Compared to the dynamic eviction policy, the performance gain from the hybrid policy is more notable when the window buffer size is smaller. These characteristics are observed when compared against both static and dynamic eviction policies because dynamic information provides more accurate insights into node access patterns when there is adequate cache size to capture temporal locality. Thus, the performance of the dynamic eviction policy heavily depends on both the window buffer size and cache capacity. The hybrid eviction policy leverages both static and dynamic information, resulting in a minimized performance decline with smaller window buffer sizes.

\begin{figure*}[ht]
    \centering
    \includegraphics[width=\textwidth]{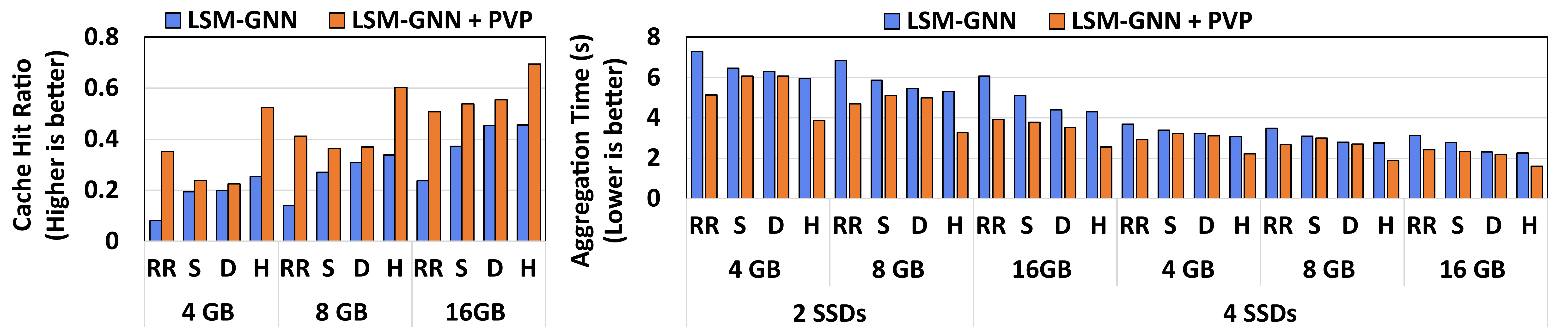}
    \caption{ Cache hit ratio and feature aggregation performance comparison between PVP and the baseline. \pname{} achieves a higher hit ratio by capturing the reusable evicted cache-lines in the Vicitm-buffer, especially with the hybrid eviction policy.}
    \label{fig:eval_pvp}
\end{figure*}

These results demonstrate that the hybrid eviction policy effectively balances static and dynamic information according to the system configuration. Consequently, users can leverage this policy across various resource availability and GNN training pipelines, ensuring optimal performance regardless of the constraints.

\subsection{Preemptive Victim-Buffer Prefetcher's Impact}

In this evaluation, we assess the impact of the Preemptive Victim-buffer Prefetcher (PVP) on GPU software cache performance and feature aggregation. We measured cache hit ratios and feature aggregation times across all configurations previously examined in the cache eviction policy assessment.

Figure~\ref{fig:eval_pvp} presents a comparison of hit ratios and feature aggregation throughputs for \pname{} both with and without PVP enabled. For the dynamic and hybrid eviction policy, we compared when the window buffer size is 256. Across all configurations, enabling PVP consistently enhances performance. Specifically, with the hybrid eviction policy, PVP elevates the hit ratio from 0.25, 0.33, and 0.48 to 0.52, 0.60, and 0.69 for cache sizes of 4 GB, 8 GB, and 16 GB, respectively. Additionally, feature aggregation throughput increases by 1.54$\times$, 1.53$\times$, and 1.68$\times$ for cache sizes of 4 GB, 8 GB, and 16 GB, respectively, when 2 SSDs are utilized. With 4 SSDs, the throughput improvements are 1.38$\times$, 1.46$\times$, and 1.51$\times$, correspondingly. These results demonstrate PVP's capability to efficiently prefetch evicted cache lines likely to be reused in subsequent iterations, storing them in the Victim-buffer for efficient CPU-to-GPU data transfer by a CPU thread.

The performance gains from PVP are less pronounced for the static and dynamic information-based eviction policies than for the Round-Robin and hybrid eviction policies. This discrepancy arises because these policies are more inclined to evict cache lines lacking dynamic information. Specifically, the dynamic information-based eviction policy prioritizes the eviction of cache lines anticipated to have no subsequent reuse, making these lines ineligible for transfer to the Victim-buffer. Conversely, while the static information-based eviction policy does exhibit a slightly higher performance gain from PVP compared to the dynamic eviction policy, it too prioritizes the eviction of cache lines with lower static values, which are less likely to be present in the window buffer. Therefore, the likelihood of evicting cache lines without dynamic information is lower with the Round-Robin and hybrid eviction policies, resulting in reduced performance benefits from PVP.

\begin{figure}[t]
    \centering
    \includegraphics[width=\columnwidth]{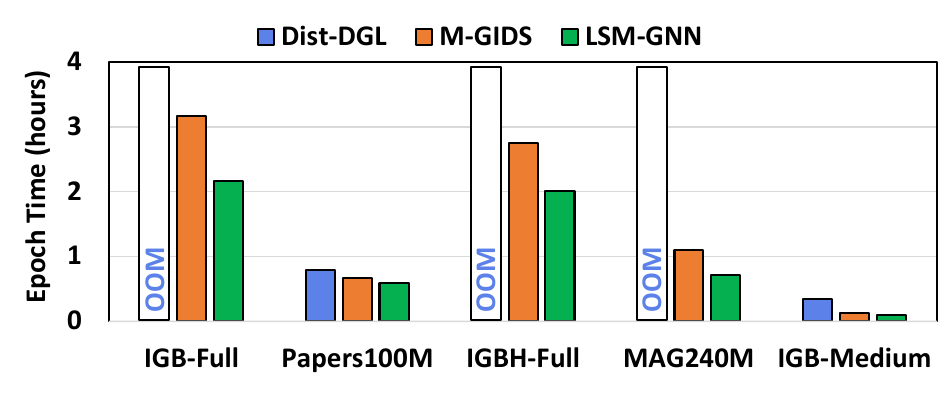}
    \vspace{-4ex}
    \caption{Epoch time comparison of LSM-GNN, M-GIDS, and Dist-DGL on homogeneous and heterogeneous graphs.}
    \label{fig:eval_overall}
\end{figure}

\subsection{Overall Performance Comparison}
Figure~\ref{fig:eval_overall} presents one epoch time for Dist-DGL, M-GIDS, and \pname{} across both homogeneous and heterogeneous graph datasets. 
\textit{Dist-DGL is executed with two 
DGX-A100  nodes each with two GPUs while M-GIDS and \pname{} are evaluated with two A100 GPUs and each GPU is connected to an SSD.}
For smaller datasets, such as ogbn-papers100M and IGB-Medium, 4 GB GPU memory is allocated for each GPU software cache in the case of M-GIDS and \pname{}. 
Due to substantial memory requirements, \textbf{Dist-DGL encounters an out-of-memory error when processing the IGB-Full, IGBH-Full, and MAG240M datasets}. Some cases trigger OOM errors due to insufficient collective GPU memory of the 4 GPUs used. 
All systems reach the same convergence properties as \pname{} or M-GIDS does not make any changes to the algorithms involved. 

\textbf{Despite the lower compute capabilities and memory capacities, M-GIDS and \pname{} in a single node with two GPUs offer superior performance over Dist-DGL baseline as shown in Figure~\ref{fig:eval_overall}.} 
\pname{} achieves up to a 1.54$\times$ speedup over M-GIDS and a 3.75$\times$ speedup over the widely used Dist-DGL. 
This superior performance of \pname{} can be attributed to novel data management strategies, including the communication layer, a hybrid eviction policy, and a Preemptive Victim-buffer Prefetcher (PVP). 
These strategies effectively alleviate storage pressure and optimize the use of GPU resources, thereby minimizing the impact of the limited storage bandwidth. The substantial speedup observed in comparison to Dist-DGL primarily results from the elimination of high network communication overheads for the distributed GNN training frameworks. 
Moreover, using fewer nodes equipped with expensive GPUs not only lowers training expenses but also significantly cuts down on power costs over time providing order of magnitude in total cost of ownership savings at scale.

%% file: related.tex
\section{RELATED WORK}
\label{sec:related}

GIDS~\cite{GIDS}, Ginex~\cite{Ginex}, and MariousGNN~\cite{MariusGNN} propose storage-based GNN training that can hide storage latency. However, due to the limited storage bandwidth and hardware resource requirement, they are not scalable in multi-GPU cases.

Previous studies~\cite{DataTiering, R1_pagerank} show PageRank can be used to estimate the frequency of accesses during node sampling, improving GPU memory utilization. 
ROC~\cite{ROC},  AliGraph~\cite{AliGraph}, PaGraph~\cite{PaGraph}, NextDoor~\cite{jangdaAcceleratingGraphSampling2021}, and Ginex~\cite{Ginex} leverage static in-memory cache to reduce communication overhead. However, unlike \pname{}'s hybrid eviction policy, these caches are solely dependent on static information. GIDS~\cite{GIDS} leverages software cache to reduce redundant accesses. However, the cache eviction policy is dependent on the number of accesses in the next iterations, increasing the contention on available cache-lines due to cache-line pinning.

Previous works~\cite{G3, dorylus, R5_sampling, BGL,Neugraph,P3, DSP, R10, GNN:I/O} propose a multi-GPU training system for large-scale GNN training. However, they require significant additional hardware resources to partition the graph across nodes or GPUs. Moreover, they are not scalable solutions due to high network communication overhead during the feature aggregation stage. Flexgraph~\cite{wangFlexGraphFlexibleEfficient2021}, Dorylus~\cite{dorylus},  ByteGNN~\cite{ByteGNN}, Pytorch-BigGraph~\cite{Pytorch-biggraph}, and Cluster-GCN~\cite{Cluster-GCN} partition graph with METIS based graph partitioning algorithm, causing a significant preprocessing overhead. 

There are also works optimizing GPU performance for the sampling stage and/or the training stage for sampling-based GNN training, including GNNadvisor~\cite{wangGNNAdvisorAdaptiveEfficient2021}, BNS-GCN~\cite{bns-gcn}, and GNNlab~\cite{GNNLab}. Our work is orthogonal to the optimizations they have proposed.


%% file: conclusion.tex
\section{CONCLUSION}

Scaling large-scale storage-based GNNs Training in multi-GPU systems is a challenging task due to the limited storage bandwidth during the feature aggregation stage. In this paper, we propose \pname{}, a large-scale storage-based multi-GPU GNN training framework, that efficiently utilizes GPU and CPU available hardware resources to scale the feature aggregation process in the multi-GPU system without changing storage configuration. \pname{} efficiently leverage GPU software caches by orchestrating each GPU cache as a shared system cache without slow system-scope operations by distributing node accesses with the communication layer. \pname{} then substantially increases the cache hit ratio by the hybrid eviction policy which exploits both static and dynamic information about the node access pattern. Finally, \pname{} leverages PVP, which temporally stores the evicted cache-line data that will be reused in the future iteration and enables high throughput feature data prefetching without increasing contention on GPU resources. All in all, \pname{} achieves up to 3.75$\times$ end-to-end training time speed up compared to the state-of-the-art  frameworks. 



%% file: Reference/GNN_backgrounds_Kun_manually_added.bib
@inproceedings{hamilton2017inductive,
author = {Hamilton, William L. and Ying, Rex and Leskovec, Jure},
title = {Inductive Representation Learning on Large Graphs},
year = {2017},
isbn = {9781510860964},
publisher = {Curran Associates Inc.},
address = {Red Hook, NY, USA},
booktitle = {Proceedings of the 31st International Conference on Neural Information Processing Systems},
pages = {1025–1035},
numpages = {11},
location = {Long Beach, California, USA},
series = {NIPS'17}
}

@ARTICLE{CNN0,
  author={Y. {LeCun} and B. {Boser} and J. S. {Denker} and D. {Henderson} and R. E. {Howard} and W. {Hubbard} and L. D. {Jackel}},
  journal={Neural Computation}, 
  title={Backpropagation Applied to Handwritten Zip Code Recognition}, 
  year={1989},
  volume={1},
  number={4},
  pages={541-551},
  doi={10.1162/neco.1989.1.4.541}}

@inproceedings{kipf2017semi,
  title={Semi-Supervised Classification with Graph Convolutional Networks},
  author={Kipf, Thomas N. and Welling, Max},
  booktitle={International Conference on Learning Representations (ICLR)},
  year={2017}
}

@inproceedings{lecunGCN,
title = "Spectral networks and locally connected networks on graphs",
author = "Joan Bruna and Wojciech Zaremba and Arthur Szlam and Yann Lecun",
year = "2014",
language = "English (US)",
booktitle = "International Conference on Learning Representations (ICLR2014), CBLS, April 2014",
}

@inproceedings{GCNPierre,
author = {Defferrard, Micha\"{e}l and Bresson, Xavier and Vandergheynst, Pierre},
title = {Convolutional Neural Networks on Graphs with Fast Localized Spectral Filtering},
year = {2016},
isbn = {9781510838819},
publisher = {Curran Associates Inc.},
address = {Red Hook, NY, USA},
booktitle = {Proceedings of the 30th International Conference on Neural Information Processing Systems},
pages = {3844–3852},
numpages = {9},
location = {Barcelona, Spain},
series = {NIPS'16}
}

@article{kipf2016variational,
  title={Variational Graph Auto-Encoders},
  author={Kipf, Thomas N and Welling, Max},
  journal={NIPS Workshop on Bayesian Deep Learning},
  year={2016}
}

@inproceedings{ying2019pinsage,
author = {Ying, Rex and He, Ruining and Chen, Kaifeng and Eksombatchai, Pong and Hamilton, William L. and Leskovec, Jure},
title = {Graph Convolutional Neural Networks for Web-Scale Recommender Systems},
year = {2018},
isbn = {9781450355520},
publisher = {Association for Computing Machinery},
address = {New York, NY, USA},
url = {https://doi.org/10.1145/3219819.3219890},
doi = {10.1145/3219819.3219890},
booktitle = {Proceedings of the 24th ACM SIGKDD International Conference on Knowledge Discovery \& Data Mining},
pages = {974–983},
numpages = {10},
location = {London, United Kingdom},
series = {KDD '18}
}

@InProceedings{pmlr-v48-niepert16, title = {Learning Convolutional Neural Networks for Graphs}, author = {Mathias Niepert and Mohamed Ahmed and Konstantin Kutzkov}, booktitle = {Proceedings of The 33rd International Conference on Machine Learning}, pages = {2014--2023}, year = {2016}, editor = {Maria Florina Balcan and Kilian Q. Weinberger}, volume = {48}, series = {Proceedings of Machine Learning Research}, address = {New York, New York, USA}, month = {20--22 Jun}, publisher = {PMLR}, url = {http://proceedings.mlr.press/v48/niepert16.html} }

@misc{linComprehensiveSurveyDistributed2022,
  title = {A {{Comprehensive Survey}} on {{Distributed Training}} of {{Graph Neural Networks}}},
  author = {Lin, Haiyang and Yan, Mingyu and Ye, Xiaochun and Fan, Dongrui and Pan, Shirui and Chen, Wenguang and Xie, Yuan},
  year = {2022},
  month = nov,
  number = {arXiv:2211.05368},
  eprint = {2211.05368},
  primaryclass = {cs},
  publisher = {{arXiv}},
  urldate = {2023-10-28},
  archiveprefix = {arxiv}
}

@inproceedings{caiDGCLEfficientCommunication2021,
  title = {{{DGCL}}: An Efficient Communication Library for Distributed {{GNN}} Training},
  shorttitle = {{{DGCL}}},
  booktitle = {Proceedings of the {{Sixteenth European Conference}} on {{Computer Systems}}},
  author = {Cai, Zhenkun and Yan, Xiao and Wu, Yidi and Ma, Kaihao and Cheng, James and Yu, Fan},
  year = {2021},
  month = apr,
  series = {{{EuroSys}} '21},
  pages = {130--144},
  publisher = {{Association for Computing Machinery}},
  address = {{New York, NY, USA}},
  doi = {10.1145/3447786.3456233},
  urldate = {2023-12-04},
  isbn = {978-1-4503-8334-9}
}

@inproceedings{wangFlexGraphFlexibleEfficient2021,
  title = {{{FlexGraph}}: A Flexible and Efficient Distributed Framework for {{GNN}} Training},
  shorttitle = {{{FlexGraph}}},
  booktitle = {Proceedings of the {{Sixteenth European Conference}} on {{Computer Systems}}},
  author = {Wang, Lei and Yin, Qiang and Tian, Chao and Yang, Jianbang and Chen, Rong and Yu, Wenyuan and Yao, Zihang and Zhou, Jingren},
  year = {2021},
  month = apr,
  series = {{{EuroSys}} '21},
  pages = {67--82},
  publisher = {{Association for Computing Machinery}},
  address = {{New York, NY, USA}},
  doi = {10.1145/3447786.3456229},
  urldate = {2023-12-04},
  isbn = {978-1-4503-8334-9}
}

@inproceedings{jouppiInDatacenterPerformanceAnalysis2017,
  title = {In-{{Datacenter Performance Analysis}} of a {{Tensor Processing Unit}}},
  booktitle = {Proceedings of the 44th {{Annual International Symposium}} on {{Computer Architecture}}},
  author = {Jouppi, Norman P. and Young, Cliff and Patil, Nishant and Patterson, David and Agrawal, Gaurav and Bajwa, Raminder and Bates, Sarah and Bhatia, Suresh and Boden, Nan and Borchers, Al and Boyle, Rick and Cantin, Pierre-luc and Chao, Clifford and Clark, Chris and Coriell, Jeremy and Daley, Mike and Dau, Matt and Dean, Jeffrey and Gelb, Ben and Ghaemmaghami, Tara Vazir and Gottipati, Rajendra and Gulland, William and Hagmann, Robert and Ho, C. Richard and Hogberg, Doug and Hu, John and Hundt, Robert and Hurt, Dan and Ibarz, Julian and Jaffey, Aaron and Jaworski, Alek and Kaplan, Alexander and Khaitan, Harshit and Killebrew, Daniel and Koch, Andy and Kumar, Naveen and Lacy, Steve and Laudon, James and Law, James and Le, Diemthu and Leary, Chris and Liu, Zhuyuan and Lucke, Kyle and Lundin, Alan and MacKean, Gordon and Maggiore, Adriana and Mahony, Maire and Miller, Kieran and Nagarajan, Rahul and Narayanaswami, Ravi and Ni, Ray and Nix, Kathy and Norrie, Thomas and Omernick, Mark and Penukonda, Narayana and Phelps, Andy and Ross, Jonathan and Ross, Matt and Salek, Amir and Samadiani, Emad and Severn, Chris and Sizikov, Gregory and Snelham, Matthew and Souter, Jed and Steinberg, Dan and Swing, Andy and Tan, Mercedes and Thorson, Gregory and Tian, Bo and Toma, Horia and Tuttle, Erick and Vasudevan, Vijay and Walter, Richard and Wang, Walter and Wilcox, Eric and Yoon, Doe Hyun},
  year = {2017},
  month = jun,
  pages = {1--12},
  publisher = {{ACM}},
  address = {{Toronto ON Canada}},
  doi = {10.1145/3079856.3080246},
  urldate = {2024-02-23},
  isbn = {978-1-4503-4892-8},
  langid = {english}
}

@misc{techpowerupGPUSpecsDatabase2024,
  title = {{{GPU Specs Database}}},
  author = {{TechPowerUp}},
  year = {2024},
  month = feb,
  urldate = {2024-02-23},
  howpublished = {https://www.techpowerup.com/gpu-specs/},
  langid = {english}
}

@misc{userbenchmarkSSDUserBenchmarks1072,
  title = {{{SSD UserBenchmarks}} - 1072 {{Solid State Drives Compared}}},
  author = {{UserBenchmark}},
  urldate = {2024-02-23},
  howpublished = {https://ssd.userbenchmark.com}
}

@misc{wikipediaTensorProcessingUnit2024,
  title = {Tensor {{Processing Unit}}},
  author = {{Wikipedia}},
  year = {2024},
  month = feb,
  urldate = {2024-02-23},
  copyright = {Creative Commons Attribution-ShareAlike License},
  url = {https://en.wikipedia.org/w/index.php?title=Tensor\_Processing\_Unit\&oldid=1209511741},
  langid = {english}
}

@article{PTXFormal,
author = {Lustig, Daniel and Sahasrabuddhe, Sameer and Giroux, Olivier},
title = {A Formal Analysis of the NVIDIA PTX Memory Consistency Model},
year = {2019},
isbn = {9781450362405},
publisher = {Association for Computing Machinery},
address = {New York, NY, USA},
url = {https://doi.org/10.1145/3297858.3304043},
doi = {10.1145/3297858.3304043},
journal = {Proc. ASPLOS},
numpages = {14},
location = {Providence, RI, USA},
}


%% file: Reference/_GENERATED_from_Kuns_Zotero.bib
@misc{epochParameterComputeData,
  title = {Parameter, {{Compute}} and {{Data Trends}} in {{Machine Learning}}},
  author = {Epoch},
  urldate = {2023-11-19},
  howpublished = {https://epochai.org/mlinputs/visualization},
  langid = {english}
}


%% file: Reference/references.bib
@misc{pytorch-direct,
      title={PyTorch-Direct: Enabling GPU Centric Data Access for Very Large Graph Neural Network Training with Irregular Accesses}, 
      author={Seung Won Min and Kun Wu and Sitao Huang and Mert Hidayetoğlu and Jinjun Xiong and Eiman Ebrahimi and Deming Chen and Wen-mei Hwu},
      year={2021},
      eprint={2101.07956},
      archivePrefix={arXiv},
      primaryClass={cs.LG}
}

@misc{GIDS,
      title={Accelerating Sampling and Aggregation Operations in GNN Frameworks with GPU Initiated Direct Storage Accesses}, 
      author={Jeongmin Brian Park and Vikram Sharma Mailthody and Zaid Qureshi and Wen-mei Hwu},
      year={2023},
      eprint={2306.16384},
      archivePrefix={arXiv},
      primaryClass={cs.DC}
}

@inproceedings{GNNLab,
author = {Yang, Jianbang and Tang, Dahai and Song, Xiaoniu and Wang, Lei and Yin, Qiang and Chen, Rong and Yu, Wenyuan and Zhou, Jingren},
title = {GNNLab: A Factored System for Sample-Based GNN Training over GPUs},
year = {2022},
isbn = {9781450391627},
publisher = {Association for Computing Machinery},
address = {New York, NY, USA},
url = {https://doi.org/10.1145/3492321.3519557},
doi = {10.1145/3492321.3519557},
booktitle = {Proceedings of the Seventeenth European Conference on Computer Systems},
pages = {417–434},
numpages = {18},
location = {Rennes, France},
series = {EuroSys '22}
}

@article{AliGraph,
author = {Zhu, Rong and Zhao, Kun and Yang, Hongxia and Lin, Wei and Zhou, Chang and Ai, Baole and Li, Yong and Zhou, Jingren},
title = {AliGraph: A Comprehensive Graph Neural Network Platform},
year = {2019},
issue_date = {August 2019},
publisher = {VLDB Endowment},
volume = {12},
number = {12},
issn = {2150-8097},
url = {https://doi.org/10.14778/3352063.3352127},
doi = {10.14778/3352063.3352127},
journal = {Proc. VLDB Endow.},
month = {aug},
pages = {2094–2105},
numpages = {12}
}

@inproceedings {BGL,
author = {Tianfeng Liu and Yangrui Chen and Dan Li and Chuan Wu and Yibo Zhu and Jun He and Yanghua Peng and Hongzheng Chen and Hongzhi Chen and Chuanxiong Guo},
title = {{BGL}: {GPU-Efficient} {GNN} Training by Optimizing Graph Data {I/O} and Preprocessing},
booktitle = {20th USENIX Symposium on Networked Systems Design and Implementation (NSDI 23)},
year = {2023},
isbn = {978-1-939133-33-5},
address = {Boston, MA},
pages = {103--118},
url = {https://www.usenix.org/conference/nsdi23/presentation/liu-tianfeng},
publisher = {USENIX Association},
month = apr,
}

@article{ByteGNN,
author = {Zheng, Chenguang and Chen, Hongzhi and Cheng, Yuxuan and Song, Zhezheng and Wu, Yifan and Li, Changji and Cheng, James and Yang, Hao and Zhang, Shuai},
title = {ByteGNN: Efficient Graph Neural Network Training at Large Scale},
year = {2022},
issue_date = {February 2022},
publisher = {VLDB Endowment},
volume = {15},
number = {6},
issn = {2150-8097},
url = {https://doi.org/10.14778/3514061.3514069},
doi = {10.14778/3514061.3514069},
journal = {Proc. VLDB Endow.},
month = {feb},
pages = {1228–1242},
numpages = {15}
}

@inproceedings{MariusGNN, 
    author = {Roger Waleffe and Jason Mohoney and Theodoros Rekatsinas and Shivaram Venkataraman},
    title = {MariusGNN: Resource-Efficient Out-of-Core Training of Graph Neural Networks}, 
    booktitle = {Proceedings of the Eighteenth European Conference on Computer Systems}, 
    year = {2023}, 
    isbn = {9781450394871}, 
    pages = {144–161},
    url = {https://doi.org/10.1145/3552326.3567501},
    publisher = {Association for Computing Machinery}
}

@inproceedings{gnn-recommendation,
author = {Fan, Wenqi and Ma, Yao and Li, Qing and He, Yuan and Zhao, Eric and Tang, Jiliang and Yin, Dawei},
title = {Graph Neural Networks for Social Recommendation},
year = {2019},
isbn = {9781450366748},
publisher = {Association for Computing Machinery},
address = {New York, NY, USA},
url = {https://doi.org/10.1145/3308558.3313488},
doi = {10.1145/3308558.3313488},
booktitle = {The World Wide Web Conference},
pages = {417–426},
numpages = {10},
location = {San Francisco, CA, USA},
series = {WWW '19}
}

@inproceedings{gnn_fraud,
author = {Liu, Zhiwei and Dou, Yingtong and Yu, Philip S. and Deng, Yutong and Peng, Hao},
title = {Alleviating the Inconsistency Problem of Applying Graph Neural Network to Fraud Detection},
year = {2020},
isbn = {9781450380164},
publisher = {Association for Computing Machinery},
address = {New York, NY, USA},
url = {https://doi.org/10.1145/3397271.3401253},
doi = {10.1145/3397271.3401253},
booktitle = {Proceedings of the 43rd International ACM SIGIR Conference on Research and Development in Information Retrieval},
pages = {1569–1572},
numpages = {4},
location = {Virtual Event, China},
series = {SIGIR '20}
}

@inproceedings{gnn_linkpredict,
author = {Zhang, Muhan and Chen, Yixin},
title = {Link Prediction Based on Graph Neural Networks},
year = {2018},
publisher = {Curran Associates Inc.},
address = {Red Hook, NY, USA},
booktitle = {Proceedings of the 32nd International Conference on Neural Information Processing Systems},
pages = {5171–5181},
numpages = {11},
location = {Montr\'{e}al, Canada},
series = {NIPS'18}
}

@inproceedings{GCN,
  author = {Kipf, Thomas N. and Welling, Max},
  booktitle = {Proceedings of the 5th International Conference on Learning Representations},
  location = {Palais des Congr{\`e}s Neptune, Toulon, France},
  series = {ICLR '17},
  title = {{Semi-Supervised Classification with Graph Convolutional Networks}},
  venue = {ICLR},
  year = 2017
}

@misc{GAT,
      title={Graph Attention Networks}, 
      author={Petar Veličković and Guillem Cucurull and Arantxa Casanova and Adriana Romero and Pietro Liò and Yoshua Bengio},
      year={2018},
      eprint={1710.10903},
      archivePrefix={arXiv},
      primaryClass={stat.ML}
}

@inproceedings{LazyGCN,
author = {Ramezani, Morteza and Cong, Weilin and Mahdavi, Mehrdad and Sivasubramaniam, Anand and Kandemir, Mahmut T.},
title = {GCN Meets GPU: Decoupling "When to Sample" from "How to Sample"},
year = {2020},
isbn = {9781713829546},
publisher = {Curran Associates Inc.},
address = {Red Hook, NY, USA},
booktitle = {Proceedings of the 34th International Conference on Neural Information Processing Systems},
articleno = {1552},
numpages = {11},
location = {Vancouver, BC, Canada},
series = {NIPS'20}
}

@misc{Metis,
author = {Karypis, George and Kumar, Vipin},
title = {METIS: A software package for
partitioning unstructured graphs, partitioning meshes, and computing
fill-reducing orderings of sparse matrices},
year ={1997}
}

@inproceedings{Graphsage,
author = {Hamilton, William L. and Ying, Rex and Leskovec, Jure},
title = {Inductive Representation Learning on Large Graphs},
year = {2017},
isbn = {9781510860964},
publisher = {Curran Associates Inc.},
address = {Red Hook, NY, USA},
booktitle = {Proceedings of the 31st International Conference on Neural Information Processing Systems},
pages = {1025–1035},
numpages = {11},
location = {Long Beach, California, USA},
series = {NIPS'17}
}

@inproceedings{pinner_sage,
author = {Pal, Aditya and Eksombatchai, Chantat and Zhou, Yitong and Zhao, Bo and Rosenberg, Charles and Leskovec, Jure},
title = {PinnerSage: Multi-Modal User Embedding Framework for Recommendations at Pinterest},
year = {2020},
isbn = {9781450379984},
publisher = {Association for Computing Machinery},
address = {New York, NY, USA},
url = {https://doi.org/10.1145/3394486.3403280},
doi = {10.1145/3394486.3403280},
booktitle = {Proceedings of the 26th ACM SIGKDD International Conference on Knowledge Discovery and Data Mining},
pages = {2311–2320},
numpages = {10},
location = {Virtual Event, CA, USA},
series = {KDD '20}
}

@inproceedings{fdgar,
author = {Wang, Jianyu and Wen, Rui and Wu, Chunming and Huang, Yu and Xiong, Jian},
title = {FdGars: Fraudster Detection via Graph Convolutional Networks in Online App Review System},
year = {2019},
isbn = {9781450366755},
publisher = {Association for Computing Machinery},
address = {New York, NY, USA},
url = {https://doi.org/10.1145/3308560.3316586},
doi = {10.1145/3308560.3316586},
booktitle = {Companion Proceedings of The 2019 World Wide Web Conference},
pages = {310–316},
numpages = {7},
location = {San Francisco, USA},
series = {WWW '19}
}

@misc{gnn_fraud3,
      title={Energy-based Out-of-Distribution Detection for Graph Neural Networks}, 
      author={Qitian Wu and Yiting Chen and Chenxiao Yang and Junchi Yan},
      year={2023},
      eprint={2302.02914},
      archivePrefix={arXiv},
      primaryClass={cs.LG}
}

@inproceedings{gnn_fraud4,
author = {Ye, Chang and Li, Yuchen and He, Bingsheng and Li, Zhao and Sun, Jianling},
title = {GPU-Accelerated Graph Label Propagation for Real-Time Fraud Detection},
year = {2021},
isbn = {9781450383431},
publisher = {Association for Computing Machinery},
address = {New York, NY, USA},
url = {https://doi.org/10.1145/3448016.3452774},
doi = {10.1145/3448016.3452774},
booktitle = {Proceedings of the 2021 International Conference on Management of Data},
pages = {2348–2356},
numpages = {9},
location = {Virtual Event, China},
series = {SIGMOD '21}
}

@misc{fewshot,
      title={Few-Shot Learning with Graph Neural Networks}, 
      author={Victor Garcia and Joan Bruna},
      year={2018},
      eprint={1711.04043},
      archivePrefix={arXiv},
      primaryClass={stat.ML}
}

@inproceedings{LP_system,
author = {Rossi, Andrea and Firmani, Donatella and Merialdo, Paolo and Teofili, Tommaso},
title = {Explaining Link Prediction Systems Based on Knowledge Graph Embeddings},
year = {2022},
isbn = {9781450392495},
publisher = {Association for Computing Machinery},
address = {New York, NY, USA},
url = {https://doi.org/10.1145/3514221.3517887},

booktitle = {Proceedings of the 2022 International Conference on Management of Data},
pages = {2062–2075},
numpages = {14},
location = {Philadelphia, PA, USA},
series = {SIGMOD '22}
}

@article{Ginex,
author = {Park, Yeonhong and Min, Sunhong and Lee, Jae W.},
title = {Ginex: SSD-Enabled Billion-Scale Graph Neural Network Training on a Single Machine via Provably Optimal in-Memory Caching},
year = {2022},
issue_date = {July 2022},
publisher = {VLDB Endowment},
volume = {15},
number = {11},
issn = {2150-8097},
url = {https://doi.org/10.14778/3551793.3551819},
doi = {10.14778/3551793.3551819},
journal = {Proc. VLDB Endow.},
month = {jul},
pages = {2626–2639},
numpages = {14}
}

@inproceedings {P3,
author = {Swapnil Gandhi and Anand Padmanabha Iyer},
title = {P3: Distributed Deep Graph Learning at Scale},
booktitle = {15th {USENIX} Symposium on Operating Systems Design and Implementation ({OSDI} 21)},
year = {2021},
isbn = {978-1-939133-22-9},
pages = {551--568},
url = {https://www.usenix.org/conference/osdi21/presentation/gandhi},
publisher = {{USENIX} Association},
month = jul
}

@inproceedings{bam,
author = {Qureshi, Zaid and Mailthody, Vikram Sharma and Gelado, Isaac and Min, Seungwon and Masood, Amna and Park, Jeongmin and Xiong, Jinjun and Newburn, C. J. and Vainbrand, Dmitri and Chung, I-Hsin and Garland, Michael and Dally, William and Hwu, Wen-mei},
title = {GPU-Initiated On-Demand High-Throughput Storage Access in the BaM System Architecture},
year = {2023},
isbn = {9781450399166},
publisher = {Association for Computing Machinery},
address = {New York, NY, USA},
url = {https://doi.org/10.1145/3575693.3575748},
doi = {10.1145/3575693.3575748},

booktitle = {Proceedings of the 28th ACM International Conference on Architectural Support for Programming Languages and Operating Systems, Volume 2},
pages = {325–339},
numpages = {15},
location = {Vancouver, BC, Canada},
series = {ASPLOS 2023}
}

@misc{IGB,
      title={IGB: Addressing The Gaps In Labeling, Features, Heterogeneity, and Size of Public Graph Datasets for Deep Learning Research}, 
      author={Arpandeep Khatua and Vikram Sharma Mailthody and Bhagyashree Taleka and Tengfei Ma and Xiang Song and Wen-mei Hwu},
      year={2023},
      eprint={2302.13522},
      archivePrefix={arXiv},
      primaryClass={cs.LG}
}

@inproceedings{DataTiering,
author = {Min, Seung Won and Wu, Kun and Hidayetoglu, Mert and Xiong, Jinjun and Song, Xiang and Hwu, Wen-mei},
title = {Graph Neural Network Training and Data Tiering},
year = {2022},
isbn = {9781450393850},
publisher = {Association for Computing Machinery},
address = {New York, NY, USA},
url = {https://doi.org/10.1145/3534678.3539038},
doi = {10.1145/3534678.3539038},
booktitle = {Proceedings of the 28th ACM SIGKDD Conference on Knowledge Discovery and Data Mining},
pages = {3555–3565},
numpages = {11},
location = {Washington DC, USA},
series = {KDD '22}
}

@Inproceedings{dgl,
  author = {Da Zheng and Chao Ma and Minjie Wang and Jinjing Zhou and Qidong Su and Xiang Song and Quan Gan and Zheng Zhang and George Karypis},
  title =  {DistDGL: Distributed Graph Neural Network Training for Billion-Scale Graphs.},
 booktitle={2020 IEEE/ACM 10th Workshop on Irregular Applications: Architectures and Algorithms (IA3)}, year={2020},
  volume={},
  number={},
  pages={36-44},
  doi={10.1109/IA351965.2020.00011}}

@misc{pyg,
      title={Fast Graph Representation Learning with PyTorch Geometric}, 
      author={Matthias Fey and Jan Eric Lenssen},
      year={2019},
      eprint={1903.02428},
      archivePrefix={arXiv},
      primaryClass={cs.LG}
}

@article{Spektral,
author = {Grattarola, Daniele and Alippi, Cesare},
title = {Graph Neural Networks in TensorFlow and Keras with Spektral [Application Notes]},
year = {2021},
issue_date = {Feb. 2021},
publisher = {IEEE Press},
volume = {16},
number = {1},
issn = {1556-603X},
url = {https://doi.org/10.1109/MCI.2020.3039072},
doi = {10.1109/MCI.2020.3039072},
journal = {Comp. Intell. Mag.},
month = {feb},
pages = {99–106},
numpages = {8}
}

@article{ROC,
  title={Improving the accuracy, scalability, and performance of graph neural networks with roc},
  author={Jia, Zhihao and Lin, Sina and Gao, Mingyu and Zaharia, Matei and Aiken, Alex},
  journal={Proceedings of Machine Learning and Systems},
  volume={2},
  pages={187--198},
  year={2020}
}

@inproceedings{DSP,
author = {Cai, Zhenkun and Zhou, Qihui and Yan, Xiao and Zheng, Da and Song, Xiang and Zheng, Chenguang and Cheng, James and Karypis, George},
title = {DSP: Efficient GNN Training with Multiple GPUs},
year = {2023},
isbn = {9798400700156},
booktitle = {Proceedings of the 28th ACM SIGPLAN Annual Symposium on Principles and Practice of Parallel Programming},
pages = {392–404},
numpages = {13},
location = {Montreal, QC, Canada},
series = {PPoPP '23}
}

@inproceedings{Neugraph,
author = {Ma, Lingxiao and Yang, Zhi and Miao, Youshan and Xue, Jilong and Wu, Ming and Zhou, Lidong and Dai, Yafei},
title = {Neugraph: Parallel Deep Neural Network Computation on Large Graphs},
year = {2019},
isbn = {9781939133038},
publisher = {USENIX Association},
address = {USA},
booktitle = {Proceedings of the 2019 USENIX Conference on Usenix Annual Technical Conference},
pages = {443–457},
numpages = {15},
location = {Renton, WA, USA},
series = {USENIX ATC '19}
}

@inproceedings{PaGraph,
author = {Lin, Zhiqi and Li, Cheng and Miao, Youshan and Liu, Yunxin and Xu, Yinlong},
title = {PaGraph: Scaling GNN Training on Large Graphs via Computation-Aware Caching},
year = {2020},
isbn = {9781450381376},
publisher = {Association for Computing Machinery},
address = {New York, NY, USA},
url = {https://doi.org/10.1145/3419111.3421281},
doi = {10.1145/3419111.3421281},
booktitle = {Proceedings of the 11th ACM Symposium on Cloud Computing},
pages = {401–415},
numpages = {15},
location = {Virtual Event, USA},
series = {SoCC '20}
}

@misc{ogbn_paper,
  title = {Open {{Graph Benchmark}}: {{Datasets}} for {{Machine Learning}} on {{Graphs}}},
  shorttitle = {Open {{Graph Benchmark}}},
  author = {Hu, Weihua and Fey, Matthias and Zitnik, Marinka and Dong, Yuxiao and Ren, Hongyu and Liu, Bowen and Catasta, Michele and Leskovec, Jure},
  year = {2021},
  month = feb,
  number = {arXiv:2005.00687},
  eprint = {2005.00687},
  eprinttype = {arxiv},
  primaryclass = {cs, stat},
  publisher = {{arXiv}},
  archiveprefix = {arXiv},
  howpublished = {\url{http://arxiv.org/abs/2005.00687}}
}

@misc{MAG,
  title = {Microsoft {{Academic Graph}}},
  author = {{Microsoft}},
  journal = {Microsoft Research},
  urldate = {2024-02-23},
  howpublished = {https://www.microsoft.com/en-us/research/project/microsoft-academic-graph/},
  langid = {american}
}

@misc{distdgl,
      title={DistDGL: Distributed Graph Neural Network Training for Billion-Scale Graphs}, 
      author={Da Zheng and Chao Ma and Minjie Wang and Jinjing Zhou and Qidong Su and Xiang Song and Quan Gan and Zheng Zhang and George Karypis},
      year={2021},
      eprint={2010.05337},
      archivePrefix={arXiv},
      primaryClass={cs.LG}
}

@inproceedings {dorylus,
author = {John Thorpe and Yifan Qiao and Jonathan Eyolfson and Shen Teng and Guanzhou Hu and Zhihao Jia and Jinliang Wei and Keval Vora and Ravi Netravali and Miryung Kim and Guoqing Harry Xu},
title = {Dorylus: Affordable, Scalable, and Accurate {GNN} Training with Distributed {CPU} Servers and Serverless Threads},
booktitle = {15th USENIX Symposium on Operating Systems Design and Implementation (OSDI 21)},
year = {2021},
isbn = {978-1-939133-22-9},
pages = {495--514},
url = {https://www.usenix.org/conference/osdi21/presentation/thorpe},
publisher = {USENIX Association},
month = jul
}

@inproceedings{imagenet,
 author = {Krizhevsky, Alex and Sutskever, Ilya and Hinton, Geoffrey E},
 booktitle = {Advances in Neural Information Processing Systems},
 editor = {F. Pereira and C.J. Burges and L. Bottou and K.Q. Weinberger},
 pages = {},
 publisher = {Curran Associates, Inc.},
 title = {ImageNet Classification with Deep Convolutional Neural Networks},
 url = {https://proceedings.neurips.cc/paper_files/paper/2012/file/c399862d3b9d6b76c8436e924a68c45b-Paper.pdf},
 volume = {25},
 year = {2012}
}

@article{model_parall1,
title = {A Hitchhiker’s Guide On Distributed Training Of Deep Neural Networks},
journal = {Journal of Parallel and Distributed Computing},
volume = {137},
pages = {65-76},
year = {2020},
issn = {0743-7315},
doi = {https://doi.org/10.1016/j.jpdc.2019.10.004},
url = {https://www.sciencedirect.com/science/article/pii/S0743731518308712},
author = {Karanbir Singh Chahal and Manraj Singh Grover and Kuntal Dey and Rajiv Ratn Shah},
}

@misc{DDP,
  title = {Getting {{Started}} with {{Distributed Data Parallel}} --- {{PyTorch Tutorials}} 2.2.1+cu121 Documentation},
  author = {{Shen Li}},
  year = {2024},
  month = feb,
  urldate = {2024-03-25},
  howpublished = {https://pytorch.org/tutorials/intermediate/ddp\_tutorial.html}
}

@misc{jangdaAcceleratingGraphSampling2021,
  title = {Accelerating {{Graph Sampling}} for {{Graph Machine Learning}} Using {{GPUs}}},
  author = {Jangda, Abhinav and Polisetty, Sandeep and Guha, Arjun and Serafini, Marco},
  year = {2021},
  month = may,
  number = {arXiv:2009.06693},
  eprint = {2009.06693},
  primaryclass = {cs},
  publisher = {{arXiv}},
  urldate = {2024-03-01},
  archiveprefix = {arxiv}
}

@inproceedings{wangGNNAdvisorAdaptiveEfficient2021,
  title = {{{GNNAdvisor}}: {{An Adaptive}} and {{Efficient Runtime System}} for {{GNN Acceleration}} on {{GPUs}}},
  booktitle = {The 15th {{USENIX Symposium}} on {{Operating Systems Design}} and {{Implementation}} ({{OSDI}} '21)},
  author = {Wang, Yuke and Feng, Boyuan and Li, Gushu and Li, Shuangchen and Deng, Lei and Xie, Yuan and Ding, Yufei},
  year = {2021},
  month = jul,
  langid = {english}
}

@inproceedings{R1_pagerank,
author = {Bojchevski, Aleksandar and Gasteiger, Johannes and Perozzi, Bryan and Kapoor, Amol and Blais, Martin and R\'{o}zemberczki, Benedek and Lukasik, Michal and G\"{u}nnemann, Stephan},
title = {Scaling Graph Neural Networks with Approximate PageRank},
year = {2020},
isbn = {9781450379984},
publisher = {Association for Computing Machinery},
address = {New York, NY, USA},
url = {https://doi.org/10.1145/3394486.3403296},
doi = {10.1145/3394486.3403296},
booktitle = {Proceedings of the 26th ACM SIGKDD International Conference on Knowledge Discovery \& Data Mining},
pages = {2464–2473},
numpages = {10},
location = {Virtual Event, CA, USA},
series = {KDD '20}
}

@inproceedings{R5_sampling,
 author = {Kaler, Tim and Stathas, Nickolas and Ouyang, Anne and Iliopoulos, Alexandros-Stavros and Schardl, Tao and Leiserson, Charles E. and Chen, Jie},
 booktitle = {Proceedings of Machine Learning and Systems},
 editor = {D. Marculescu and Y. Chi and C. Wu},
 pages = {172--189},
 title = {Accelerating Training and Inference of Graph Neural Networks with Fast Sampling and Pipelining},
 url = {https://proceedings.mlsys.org/paper_files/paper/2022/file/afacc5db3e0e85b446e6c7727cd7dca5-Paper.pdf},
 volume = {4},
 year = {2022}
}

@inproceedings{R10,
author = {Song, Shihui and Jiang, Peng},
title = {Rethinking graph data placement for graph neural network training on multiple GPUs},
year = {2022},
isbn = {9781450392815},
publisher = {Association for Computing Machinery},
address = {New York, NY, USA},
url = {https://doi.org/10.1145/3524059.3532384},
doi = {10.1145/3524059.3532384},
booktitle = {Proceedings of the 36th ACM International Conference on Supercomputing},
articleno = {39},
numpages = {10},
location = {Virtual Event},
series = {ICS '22}
}

@inproceedings{bns-gcn,
 author = {Wan, Cheng and Li, Youjie and Li, Ang and Kim, Nam Sung and Lin, Yingyan},
 booktitle = {Proceedings of Machine Learning and Systems},
 editor = {D. Marculescu and Y. Chi and C. Wu},
 pages = {673--693},
 title = {BNS-GCN: Efficient Full-Graph Training of Graph Convolutional Networks with Partition-Parallelism and Random Boundary Node Sampling},
 url = {https://proceedings.mlsys.org/paper_files/paper/2022/file/676638b91bc90529e09b22e58abb01d6-Paper.pdf},
 volume = {4},
 year = {2022}
}

@inproceedings{Cluster-GCN,
author = {Chiang, Wei-Lin and Liu, Xuanqing and Si, Si and Li, Yang and Bengio, Samy and Hsieh, Cho-Jui},
title = {Cluster-GCN: An Efficient Algorithm for Training Deep and Large Graph Convolutional Networks},
year = {2019},
isbn = {9781450362016},
publisher = {Association for Computing Machinery},
address = {New York, NY, USA},
url = {https://doi.org/10.1145/3292500.3330925},
doi = {10.1145/3292500.3330925},
booktitle = {Proceedings of the 25th ACM SIGKDD International Conference on Knowledge Discovery \& Data Mining},
pages = {257–266},
numpages = {10},
location = {Anchorage, AK, USA},
series = {KDD '19}
}

@inproceedings{Pytorch-biggraph,
 author = {Lerer, Adam and Wu, Ledell and Shen, Jiajun and Lacroix, Timothee and Wehrstedt, Luca and Bose, Abhijit and Peysakhovich, Alex},
 booktitle = {Proceedings of Machine Learning and Systems},
 editor = {A. Talwalkar and V. Smith and M. Zaharia},
 pages = {120--131},
 title = {Pytorch-BigGraph: A Large Scale Graph Embedding System},
 url = {https://proceedings.mlsys.org/paper_files/paper/2019/file/1eb34d662b67a14e3511d0dfd78669be-Paper.pdf},
 volume = {1},
 year = {2019}
}

@article{G3,
author = {Wan, Xinchen and Xu, Kaiqiang and Liao, Xudong and Jin, Yilun and Chen, Kai and Jin, Xin},
title = {Scalable and Efficient Full-Graph GNN Training for Large Graphs},
year = {2023},
issue_date = {June 2023},
publisher = {Association for Computing Machinery},
address = {New York, NY, USA},
volume = {1},
number = {2},
url = {https://doi.org/10.1145/3589288},
doi = {10.1145/3589288},
month = {jun},
journal = {Proc. ACM Manag. Data},
articleno = {143},
numpages = {23}
}

@INPROCEEDINGS{GNN:I/O,
  author={Lee, Claire Songhyun and Hewes, V and Cerati, Giuseppe and Kowalkowski, Jim and Aurisano, Adam and Agrawal, Ankit and Choudhary, Alok and Liao, Wei-Keng},
  booktitle={2023 IEEE/ACM 23rd International Symposium on Cluster, Cloud and Internet Computing (CCGrid)}, 
  title={A Case Study of Data Management Challenges Presented in Large-Scale Machine Learning Workflows}, 
  year={2023},
  volume={},
  number={},
  pages={71-81},
  doi={10.1109/CCGrid57682.2023.00017}
}
